\documentclass{aa}  

\usepackage{graphicx}
\usepackage{txfonts}
\usepackage{natbib}

\begin{document}

   \title{Geometry of the X-ray source 1H 0707--495}

   \author{Micha{\l} Szanecki\inst{1} \and Andrzej Nied{\'z}wiecki\inst{2} \and Chris Done\inst{3} \and {\L}ukasz Klepczarek\inst{2} \and Piotr Lubi{\'n}ski\inst{4} \and Misaki~Mizumoto\inst{5,6,3}
          }

   \institute{Nicolaus Copernicus Astronomical Center, Polish Academy of Sciences, Bartycka 18, PL-00-716 Warszawa, Poland\\
              \email{mitsza@camk.edu.pl}
	\and
			Faculty of Physics and Applied Informatics, {\L}{\'o}d{\'z} University, Pomorska 149/153, PL-90-236 {\L}{\'o}d{\'z}, Poland
	\and	
			Centre for Extragalactic Astronomy, Department of Physics, University of Durham, South Road, Durham, DH1 3LE, UK
	\and
			Institute of Physics, University of Zielona G\'{o}ra, Licealna 9, PL-65-417 Zielona G\'{o}ra, Poland
	\and
			Hakubi Center, Kyoto University, Yoshida-honmachi, Sakyo-ku, Kyoto, 606-8501, Japan
	\and
	        Department of Astronomy, Kyoto University, Kitashirakawa-oiwakecho, Sakyo-ku, Kyoto 606-8502, Japan
}

   \date{Received; accepted}

 
  \abstract
   {}
   {We investigate constraints for the size and location of the X-ray source in 1H 0707--495 determined from the shape of the relativistically smeared reflection from the accretion disc.}
   {We develop a new code to model an  extended X-ray source and we apply it to all archival {\it XMM} observations of 1H 0707--495.}
   {Contrary to Wilkins et al.\ we find that the relativistic reflection in this source is not consistent with an extended uniform corona. Instead, we find that the X-ray source must be very compact, with the size of at most a gravitational radius, and located at most at a few gravitational radii from the black hole horizon. A uniform extended corona does indeed produce an emissivity which is like a twice broken power law, but the inner emissivity is fixed by the source geometry rather than being a free parameter. In 1H0707--495, reflection from the inner disc is much stronger than expected for a uniformly extended source. 
   Including the effect of ionised absorption from a wind does not change this conclusion, but including scattered emission (and more complex absorption) from the wind can dramatically change the reflection parameters. 
  }
   {}


   \maketitle
%

\section{Introduction}

1H 0707--495 is a narrow line Seyfert 1 galaxy well known for a sharp drop in its spectrum around 7 keV \citep{2002MNRAS.329L...1B}. This spectral feature  can be imprinted either by an  absorbing material on our line of sight  \citep[e.g.][]{2004PASJ...56L...9T,2007MNRAS.374L..15D,2016MNRAS.461.3954H,2019MNRAS.482.5316M}, or by irradiation of an accretion disc \citep[e.g.][]{2009Natur.459..540F,2010MNRAS.401.2419Z,2012MNRAS.422.1914D}. Here we consider the latter and we study constraints on the size and location of the X-ray source which can be derived under this spectral interpretation. 
   
The relativistic reflection spectroscopy is an important tool for constraining geometry of accretion flows in their inner parts \citep[e.g.][]{1995Natur.375..659T,2010SSRv..157..167F,2013mams.book.....B}. Two  approximations are commonly used for such studies to simplify the model parametrization and computations.
First, in the so called lamp-post geometry, irradiation of the disc surface by a point-like source is assumed \citep[][and many later works]{1996MNRAS.282L..53M}. 
This allows to estimate the distance of the X-ray source, but not its size.
Secondly, even more often, a phenomenological radial emissivity, typically a power-law, is assumed, without any explicit assumption about the X-ray source. Here, departures of the estimated emissivity from the dissipation profile of a Keplerian disc indicate that effects of the source-to-disc radiation transfer may be involved in shaping the reflection spectrum. Then, the fitted emissivity can be used to infer the source geometry.

Such an approach was proposed by \citet[][hereafter WF12]{2012MNRAS.424.1284W}. They
showed the apparent similarity between an empirical, double-broken power law radial emissivity and the actual emissivity of the disc irradiated by an extended source, and used this to 
find the time averaged size of the X-ray source in 1H 0707--495 of about 30 gravitational radii. 
Then, \citet[][hereafter W14]{2014MNRAS.443.2746W} used spectra binned by the  flux to show that the
inferred outer break radius of the emissivity decreases with flux, and again compared these emissivities with their extended source illumination to claim that the size of the extended 
X-ray source decreases with decreasing source luminosity. 

In order to gauge the accuracy of their approach, we develop here a similar relativistic reflection model with an extended X-ray source. However, rather than limiting our study to the qualitative comparison of this model with empirical emissivities (as WF12 and W14 did) we use it directly for spectral description of 1H 0707--495. 
A part of this work addresses WF12 and W14, therefore, we adopt their data reduction method and assumptions about the X-ray source, and we use the same rest-frame reflection model rather than the newer ones now available. 

We also discuss in detail  the reflection strength, which is the normalization of the observed reflection with respect to the direct spectral component from the X-ray source. This quantity is strictly determined for a given geometry \citep[see, e.g.,][]{2014MNRAS.444L.100D} and, then, it gives a crucial test for the self-consistency of a model. However, this important constraint is often neglected in the X-ray data analysis. This concerns, in particular, all relativistic reflection models of 1H 0707--495 noted above. 

Finally, evidence for a blueshifted photoionised outflow in 1H 0707--495 was reported by \citet{2012MNRAS.422.1914D}, \citet{2016MNRAS.461.3954H} and \citet{2018MNRAS.481..947K}. We take it into account and investigate how the presence of such a blueshifted absorber affects the fitted parameters of our reflection model.

\section{Data reduction}

We consider 15 {\it XMM-Newton} \citep{2001A&A...365L...1J} observations of 1H 0707--495 
between 2000 and 2011 with an exposure time longer than 10 ks,  see e.g.\ table 1 in \citet{2016MNRAS.461.3954H}.
The 15 spectra from the EPIC pn detector \citep{2001A&A...365L..18S} were extracted using the {\it XMM-Newton} Science Analysis Software (SAS) 
version 17.0.0, and the Current Calibration File released on 29-Mar-2019. For 
each observation 
the data were reduced separately. We applied the standard selection criteria 
with the \texttt{PATTERN $\leq 4$} condition and excluding periods with the 10--12 keV background rate above 0.4 counts per second. 
The source spectra were extracted for a circular region
having 60 arcsec in diameter. For the background spectra we have found optimal
circular region using the \texttt{ebkgreg} tool, which was typically chosen to be a region of 
about 120 arcsec in diameter.
The response files for each spectrum were generated with the \texttt{rmfgen} and \texttt{arfgen} 
tools.

The observation performed in January 2011 (Obs.\ ID 0554710801) caught the
source in an extremely low state \citep{2012MNRAS.419..116F} and is treated separately here; we refer to it as the very low (VL) state. The remaining fourteen observations  
1

are used together to build the spectra of three non-overlapping count rate intervals,  with $ < 4$, 4 -- 6 and 6 -- 10 counts/s (in the 0.2 -- 10
keV energy range), which we refer to as the low (L), medium (M) and high (H) state,
respectively. This selection of flux-resolved spectra is similar to that used in W14 and it is based on the assumption that the state of the X-ray source is determined by the X-ray flux. Spectral differences between these spectra could be then used to verify whether systematic changes of the source geometry correspond to the change of the observed flux.

Light curves of 1H 0707--495 with the time bin equal to 100 s were extracted 
with the same size of the source and background regions as the spectra and 
corrected with the \texttt{epiclccorr} tool. Using the light curves we created good time 
intervals (GTI) files for each of the 14 observations, applying the above count rate criteria. The count rate selected spectra were extracted using
these GTI files, with appropriate response files. Finally, the spectra for each
count rate range were summed with the \texttt{epicspeccombine} tool. 
The resulting spectra  have total exposure times of 64.5 (VL), 313.5 (L), 318 (M) and 237.8 (H) ks.

Our selection criteria for flux-resolved spectra are almost the same as those applied in W14, except for the energy range used for spectral selection, which in our case is slightly larger than the 0.3 -- 10 keV range used in that work. Our extracted light curves are then essentially the same as those shown in Figure 1 of W14, except for a slightly larger (by about 0.5 counts/s) amplitude of ours.

Following W14 we consider the 1.1--10 keV range.
For spectral binning we adopt two approaches. 
For our basic version of the binning, referred to as binning (1),
we use the \texttt{specgroup} tool of SAS with the minimal 
signal-to-noise ratio for each channel set to 5 and the oversample parameter set 
to 3.
To address the results of W14, we use also binning (2) following their procedure, i.e.\
using the \texttt{grppha} tool of the \texttt{HEASOFT} package, with the condition of having at least 25 counts in each
spectral bin. We note large differences regarding assessments of the agreement of the considered spectral model with the spectra binned using procedures (1) and (2).

\section{Model}
\label{sect:model}

The model developed here, referred to as \texttt{reflkerr\_elp} (for 'extended lamp-post'),  closely follows the lamp-post model \texttt{reflkerr\_lp} of \citet{2019MNRAS.485.2942N}, except for taking into account the spatial extent and rotation of an X-ray source, instead of treating it as point-like and static (lamp-post). We consider a Keplerian disc in the Kerr metric, irradiated by a cylindrical corona with radius $r_{\rm c}$, located symmetrically around the black-hole rotation axis between a lower height $h_{\rm min}$ and upper height $h_{\rm max}$. The inclination angle of a distant observer is denoted by $i$. All length scales $r,h$ are in units of the gravitational radius, $R_{\rm g} = GM/c^2$. All the results presented below are for $a=0.998$. We assume that the corona is uniform, i.e.\ its rest-frame  emissivity is constant in Boyer-Lindquist coordinates. 

The finite size of the corona means that we should also include its dynamics. We assume that the corona corotates with the disc, i.e.\ the angular velocity of the corona at ($r$, $\theta$) equals the Keplerian velocity, $\Omega_{\rm K} (\rho)$, where $\rho = r \sin \theta$. For $\rho <  r_{\rm ISCO}$ we assume rigid rotation with $\Omega_{\rm K} (r_{\rm ISCO})$, where $r_{\rm ISCO}$ is the radius of the innermost stable circular orbit.

The inner radius of the disc is denoted by $r_{\rm in}$ and we assume that $r_{\rm in} \ge r_{\rm ISCO}$. 1H 0707--495 may be accreting at a rate close to the Eddington limit \citep[see e.g.][]{2016MNRAS.460.1716D}, which may mean that the region of the flow at $r < r_{\rm ISCO}$ is optically thick so it can contribute to the observed reflection \citep{1997ApJ...488..109R}. This effect is not included in all previous studies of this object. We check that this is indeed negligible in our models (all of which have extreme $a$,  and inclination angles $i \simeq (40-50) \degr$).

Our computation of the fluxes of the X-ray radiation, either reaching directly the distant observer or illuminating the disc surface, strictly follows the model of \citet[][hereafter NZ08]{2008MNRAS.386..759N}; see Appendix \ref{app:a} for technical details of our \texttt{xspec} implementation and comparison with similar computations.

To compute the observed reflected component, we apply the procedure used in the  \texttt{reflkerr\_lp} model, with  \texttt{reflionx} \citep{2005MNRAS.358..211R} used for the rest-frame reflection. In Sections \ref{sect:bpl} and \ref{sect:wind} we apply the  \texttt{reflkerr} model \citep{2019MNRAS.485.2942N}, in which we also use \texttt{reflionx}  for the rest-frame reflection (instead of  \texttt{hreflect} originally used in \texttt{reflkerr}).
In our fitting procedure we use a free normalization of the reflection component, which allows to compare our results with previous works. We scale the reflection normalization by $\mathcal{R}$, where $\mathcal{R}=1$ corresponds to the actual normalization for a given geometry. For the discussion below we also define the reflection strength, $\mathcal{R}_{\rm obs}$,  defined here as the ratio of the observed reflected and direct fluxes in the 20-30 keV range.

   \begin{figure}
   \centering
\includegraphics[width=8.5cm]{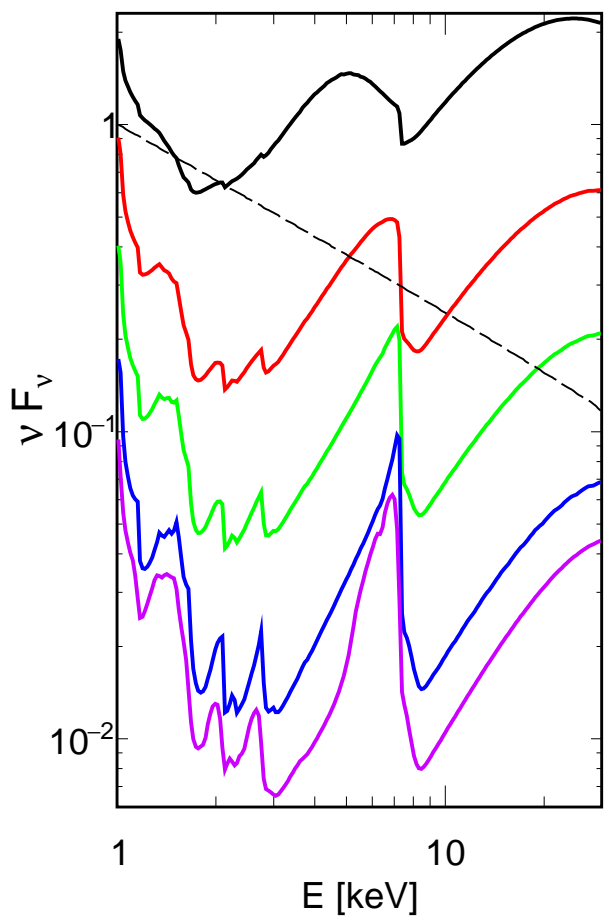}
      \caption{Illustration of changes of the reflection spectrum resulting from the contraction of the X-ray source, described by the decrease of both 
$h_{\rm max}$  and $r_{\rm c}$, for $\Gamma=2.6$, $\xi=50$, Z$_{\rm Fe}=9$, $i=50 \degr$, $\mathcal{R}=1$, $a=0.998$, $h_{\rm min}=0$ and $r_{\rm in} = r_{\rm ISCO}$. 
The reflection spectra, shown by the solid curves, correspond to $h_{\rm max}= r_{\rm c}= 1.3$ (black), 2 (red), 3 (green), 8 (blue) and 30 (magenta) from top to bottom. The primary spectrum is the same in all cases and is shown by the dashed line.
              }
         \label{size}
   \end{figure}

\begin{figure}
   \centering
\includegraphics[width=9cm]{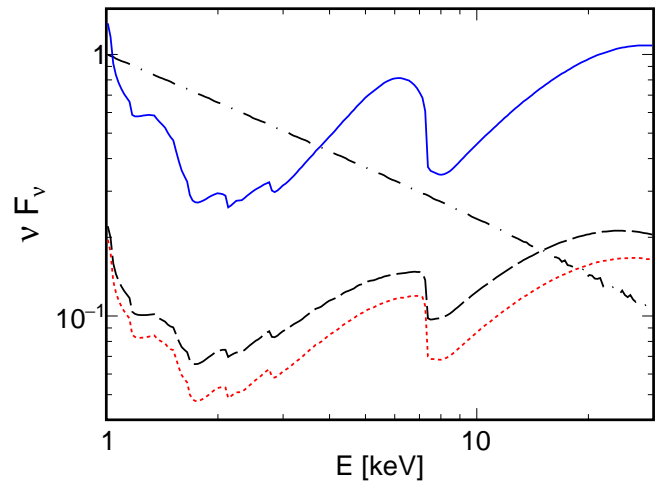}
\caption{
An example  illustration of the effect of the rotation of the X-ray source.
The dotted red curve shows the reflection spectrum computed with \texttt{reflkerr\_lp}, i.e.\  for a static point-like source, at $h=2$. The dashed black and solid blue curves show the reflection spectra computed with \texttt{reflkerr\_elp} with $r_{\rm c}=0.2$ and 1.5, respectively; in both $h_{\rm min}=h_{\rm hor}$ and $h_{\rm max}=2$. In all three cases $\Gamma=2.6$, $\xi=50$, A$_{\rm Fe}=9$, $i=50 \degr$, $a=0.998$, $r_{\rm in} = r_{\rm ISCO}$ and $\mathcal{R}=1$. The observed primary spectrum is the same in all three models and it is shown by the dot-dashed black curve. The difference between the \texttt{reflkerr\_lp} and the \texttt{reflkerr\_elp} spectra is primarily due to the azimuthal motion of the emitting plasma in the latter, with $V \simeq 0.1$c for $r_{\rm c}=0.2$ and $\simeq 0.8$c for $r_{\rm c}=1.5$.
} 
\label{fig:v}
\end{figure}

Figure \ref{size} illustrates changes of the reflection spectra corresponding to the contraction of the X-ray source in the vicinity of the black hole horizon. We  start with the magenta line showing a fairly large and thick corona, with $h_{\rm max}=r_{\rm c}=30$ and $h_{\rm min}=0$. We successively shrink $h_{\rm max}=r_{\rm c}$ to 8 (blue) 3 (green) 2 (red) and 1.3 (black). The increase in relativistic broadening for smaller sizes is clear, as is the increasing $\mathcal{R}_{\rm obs}$. The latter results from the combination of Doppler beaming and light bending. The latter effect is well known to increase the fraction of photons hitting the disc \citep{1996MNRAS.282L..53M}, however, for a static source, the related enhancement of the observed reflection is rather weak (except for large $i$, see below). If the X-ray source rotates, its radiation is Doppler beamed to outer parts of the disc. Then, the observed reflection is not so strongly dependent on $i$. At the same time, the change of the irradiation profile resulting from the source rotation leads to a considerable change of the observed spectral shape of reflection. 

Figure \ref{fig:v} illustrates the effect of azimuthal motion of the X-ray source; see also NZ08, who extensively studied the effects of the off-axis location of the X-ray source and its orbital motion. Our assumptions about co-rotation with the disc at $\rho \ge r_{\rm ISCO}$ and rigid rotation at $\rho < r_{\rm ISCO}$ appear to be natural, and are the only choices which can be justified {\it ab initio}, unlike the unknown dynamics associated with magnetic flares.  The angular velocity within  $r_{\rm ISCO}$ in this model is equal to $\Omega_{\rm K} (r_{\rm ISCO})$, giving $V \simeq 0.1$c at $\rho=0.2$, where $V$ is the velocity measured in the locally non-rotating frame \citep{1972ApJ...178..347B}.  At $h=1.5$ and $\rho=1.5$, $V \simeq 0.8$c is much larger than the Keplerian $V$ in the equatorial plane. This is because $V$ of a co-rotating source for a Kerr black hole increases with $h$ at $h \la \rho$, see section 2.2.3 in NZ08.

Figure \ref{fig:v} shows results at $i=50 \degr$, relevant for 1H0707--495. For a static point-like source at $h=2$ (the red dotted curve) $\mathcal{R}_{\rm obs}/\mathcal{R}_{\rm obs,nr} \simeq 2$, although in this case $N_{\rm irr}/N_{\rm obs} \simeq 5$ (where $\mathcal{R}_{\rm obs,nr}$ is the reflection strength for an isotropic source above a semi-infinite slab in flat space-time, and $N_{\rm irr}$ and $N_{\rm obs}$ are, respectively, the number of photons hitting the disc and observed directly from the source). This factor of 2 enhancement of the reflection strength is lower than the factor 5 enhancement of irradiating photon number compared to those observed as for a static lamp-post, the enhanced irradiation occurs mostly at $r \la h$ and the reflected radiation is directed mostly to large $i$ \cite[see e.g.][]{2018MNRAS.477.4269N}; therefore, the related increase of $\mathcal{R}_{\rm obs}$ is rather modest for $i \la 70 \degr$. 

The black dashed curve in Figure \ref{fig:v} shows a mostly vertical extended source from the horizon i.e. $h_{\rm min}=h_{\rm hor}$ up to $h_{\rm max}=2$, with very small radial extent $r_{\rm c}=0.2$. This geometry is very similar to the lamp-post for lightbending but the rotation velocity of $V \simeq 0.1$c Doppler beams the radiation more toward the outer parts of the disc, where reflection gives a stronger contribution for small $i$ and, hence, $\mathcal{R}_{\rm obs}$ is larger.

This effect is even more marked when the radial extent of the source is increased to $r_{\rm c}=1.5$ (blue solid curve), as this has $V \simeq 0.8$c.
and $\mathcal{R}_{\rm obs}/\mathcal{R}_{\rm obs,nr} \simeq 15$. The increasing distortion of the reflection spectrum, as compared to the static source, is clearly seen.

   \begin{figure}
   \centering
\includegraphics[width=8.5cm]{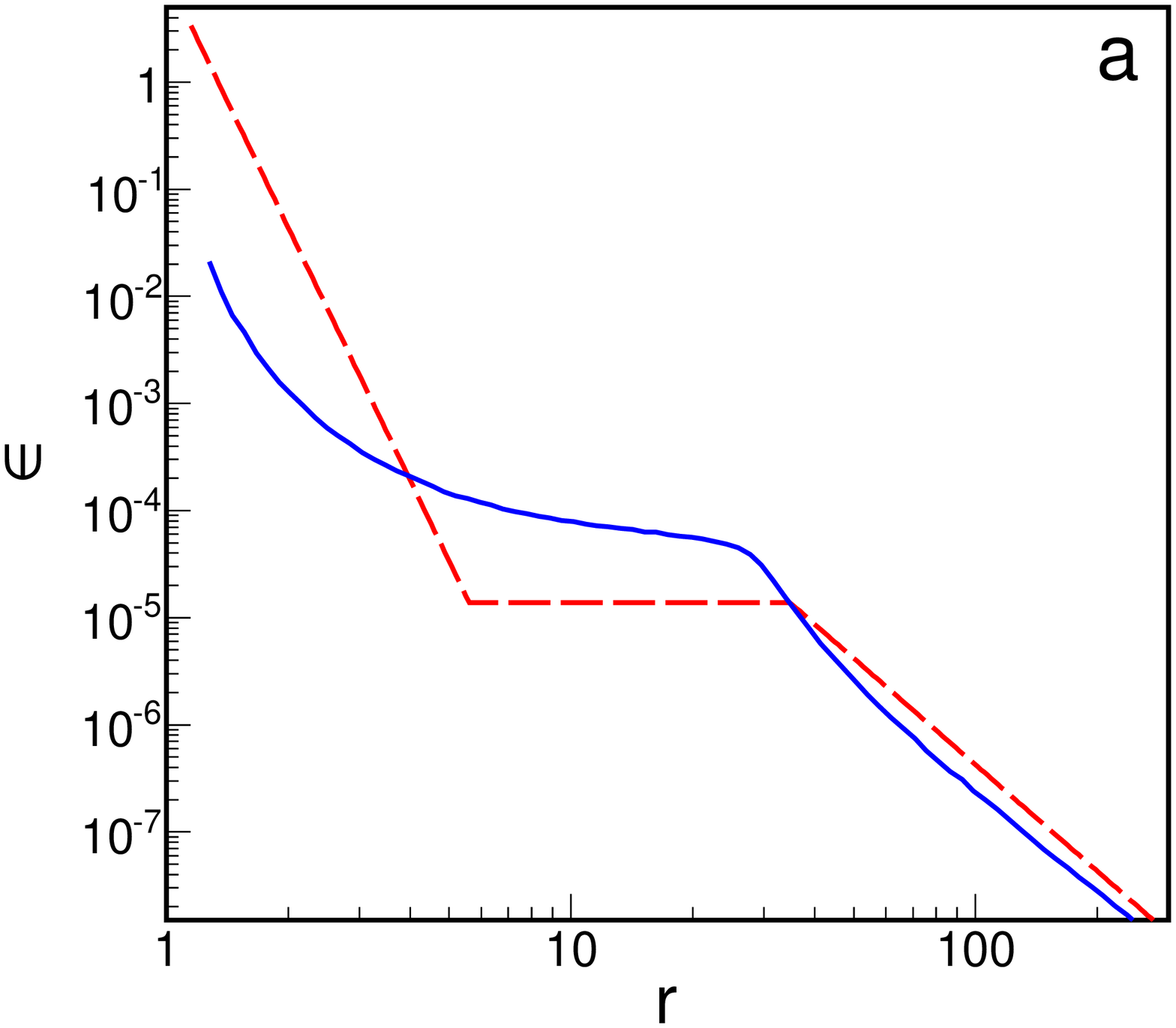}
\vspace{0.2cm}
\includegraphics[width=8.5cm]{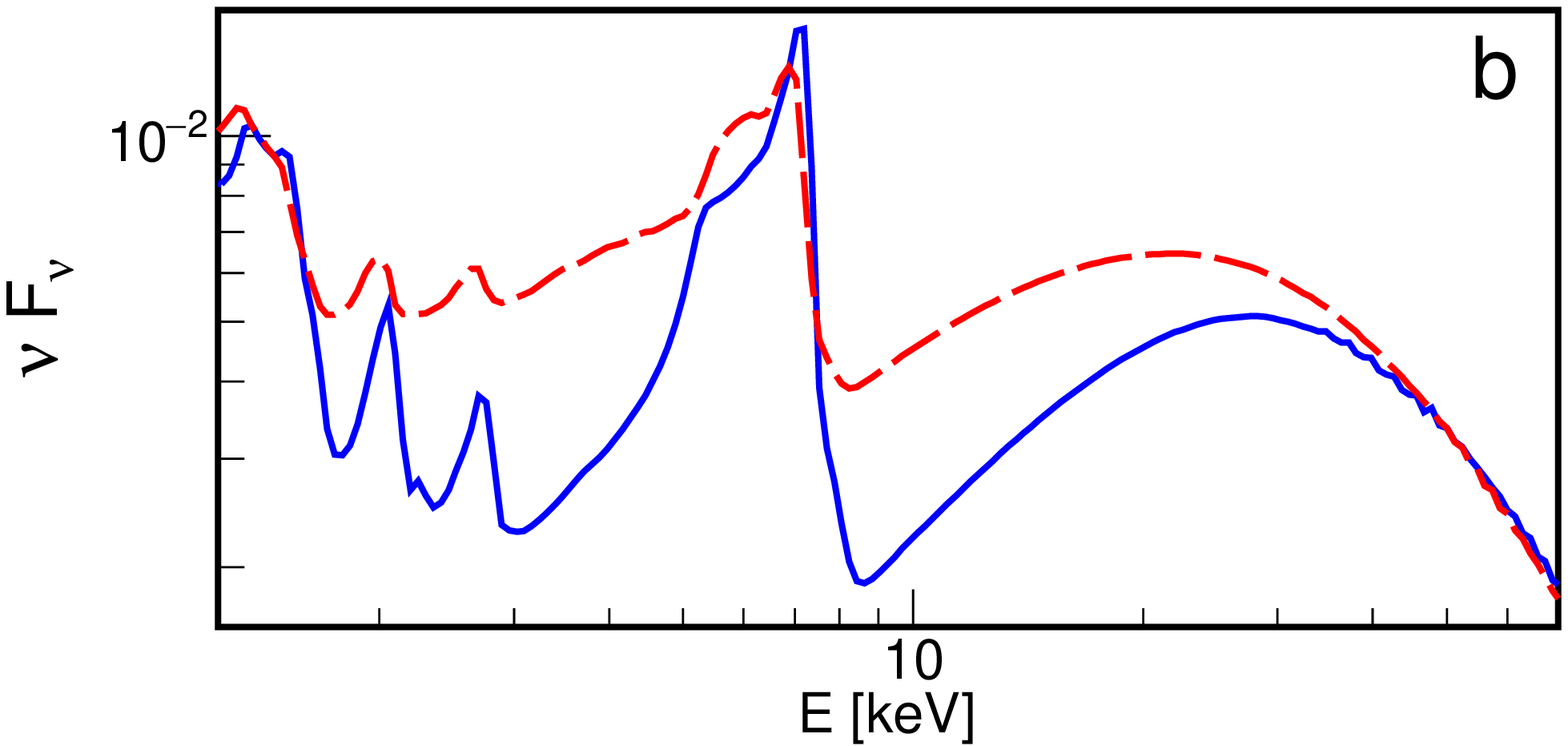}
      \caption{(a) 
The red line shows the twice-broken power law radial emissivity profile fitted to 1H0707--495 by \cite{2011MNRAS.414.1269W}, see text.
The blue line shows the emissivity profile produced by the X-ray source with $h_{\rm min}=2$, $h_{\rm max}=10$, $r_c=30$ and $\Gamma=3.1$.
(b) Reflection spectra for the emissivity profiles shown in (a), with $\Gamma=3.1$, $\xi=53$, A$_{\rm Fe}=9$, $i=54 \degr$ and $r_{\rm in} = r_{\rm ISCO}$; the difference of spectral shapes is only due to different amounts of relativistic distortion corresponding to the difference of the radial profiles.
              }
         \label{radial}
   \end{figure}

\begin{table}
\centering
\caption{The results of spectral fitting of the coronal model \texttt{reflkerr} using empirical radial emissivities. The model fitted to L, M and H assumes a twice-broken power law radial emissivity profile and $\xi = 53.44$. The model fitted to VL assumes a once-broken power law profile; $\xi$ is free. In all models  $i = 53.96^{\circ}$,  and $Z_{\rm Fe} = 8.88$, $a=0.998$, $r_{\rm in}=r_{\rm ISCO}$ and $r_{\rm out}=1000$. The $\chi^{2}/$DoF values correspond to spectral binning (1) and (2), see Section 2. The best-fit values and parameter uncertainties given in the table correspond to binning (1); the differences in the best-fit values obtained with binning (2) are insignificant.}
\begin{tabular}{ lllll  }
 \multicolumn{5}{c}{Model: \texttt{reflkerr}} \\
 \hline
    			& VL		& L 			& M 		& H \\
 \hline
 $q_1$ 	  		& $-0.8^{+1.4}_{-\infty}$ 	& 8.9$^{+0.5}_{-0.6}$ 	& 8.4$^{+0.5}_{-0.4}$ 	& 6.6$^{+1.1}_{-0.9}$ 	\\[0.1cm]
 $q_2$		  	& $7.2^{+2.1}_{-1.1}$ 	& 0.0$^{+0.9}_{-0}$ 	& 0.1$^{+0.7}_{-0.1}$	& 1.4$^{+0.5}_{-0.8}$ 	\\[0.1cm]
 $q_3$ 			& - 			& 3.0$^{+0.5}_{-0.4}$ 	& 4.0$^{+0}_{-0.7}$	& 4.0$^{+0}_{-0.6}$ 	\\[0.1cm]
 $r_{\rm br,1}$ 	& $2.3^{+0.3}_{-0.3}$	& 4.5$^{+0.5}_{-0.6}$   & 5.0$^{+0}_{-0.51}$ 	& 4.7$^{+0.3}_{-0.7}$ 	\\[0.1cm]
 $r_{\rm br,2}$ 	& - 			& 18.5$^{+9.7}_{-5.6}$  & 28.2$^{+5.7}_{-5.7}$	& 32.0$^{+3.0}_{-9.1}$ 	\\[0.1cm]
 $\Gamma$		& $2.69^{+0.04}_{-0.16}$& 3.20$^{+0.07}_{-0.07}$& 3.10$^{+0.06}_{-0.03}$& 3.00$^{+0.07}_{-0.03}$\\[0.1cm]
 $\xi$			& $10^{+2.0}_{-0}$ 	& $53.44^{(f)}$ 	& $53.44^{(f)}$ 	& $53.44^{(f)}$ 	\\[0.1cm]
 $\mathcal{R}$		& $> 750$ 		& $88.2^{+20.4}_{-18.3}$& $29.1^{+5.1}_{-3.3}$ 	& 11.4$^{+3.1}_{-1.3}$	\\[0.1cm]
 \hline
 & \multicolumn{4}{c}{$\chi^{2}/$DoF} \\[0.1cm]
 \hline

  (1) 	&  119/86  	& 290/132	  	& 293/141 	    	& 347/140\\
  (2)	&  173/127  	& 836/688	  	& 1089/932 	    	& 1123/950\\

 \hline
 \end{tabular}\\
\tablefoot{{\it (f)} denotes a fixed parameter. Following W14, for L, M and H the fitted ranges of $q_1$, $q_2$, $q_3$, $r_{\rm br,1}$ and $r_{\rm br,2}$ are constrained to $[5,\,10]$, $[0,\,2]$, $[2,\,4]$, $[3,\,5]$ and $[5,\,35]$, respectively. The uncertainties are given for 90\% confidence level. The uncertainties given as '+0' or '-0' correspond to the limit of the considered range of a parameter. $\xi$ is given in the unit of ${\rm erg\; cm\; s}^{-1}$.}
\label{tab:bpl}
\end{table}

Concluding the above discussion we note that significant changes of the angular distribution of radiation at the X-ray source (in particular, those related to the motion of the emitting plasma) may significantly change $\mathcal{R}_{\rm obs}$. However, they also lead to significant changes of the observed shape of the reflected spectrum (due to the change of the radial irradiation profile).  Then, treating the reflection strength as a free parameter is unphysical and we again note that we scale it by $\mathcal{R}$ only to compare our results with earlier works. Models with $\mathcal{R} \gg 1$, such as found in Sections \ref{sect:bpl} and \ref{sect:refl}, are clearly unphysical.

\section{Comparison with empirical radial emissivity}
\label{sect:comp}

The red curve in Figure \ref{radial}(a) shows the twice-broken power law emissivity fitted to average spectra of 1H0707--495 by \citet{2011MNRAS.414.1269W}, with the inner, middle and outer indices of $q_1=7.8$, $q_2=0$ and $q_3=3.3$, respectively, and the breaking radii of $r_{\rm br,1}=5.6$ and $r_{\rm br,2}=35$. 
WF12 argued that this inferred emissivity profile is consistent with that produced by an X-ray source with $r_{\rm c} \simeq r_{\rm br,2}$
(specifically, with $h_{\rm min}=2$, $h_{\rm max}=10$ and $r_{\rm c}=30$). We use our code to explicitly 
calculate the emissivity of such an extended corona
(blue curve in Figure \ref{radial}a). Our 
assumptions concerning the X-ray source are exactly the same as those of WF12,
and indeed our radial profile agrees with theirs (Figure 15(b) in WF12). However, it differs substantially from the empirical profile of \cite{2011MNRAS.414.1269W}. 
The empirical profile fitted to 1H0707--495 indicates a major contribution of radiation reflected from $r < 5$, which suffers strong redshift. However, while the irradiation profile produced by an extended source does steepen at 
$r \la 5$ (due to the blueshift of photons arriving at this region from X-rays produced in the corona at larger distances), this effect is  too weak to reproduce
the extremely steep (with $q \simeq 8$) inner part of the empirical profile. The contribution of reflection arising from $r < 5$ is only 15 per cent for the actual profile of an extended corona irradiating the disc, whereas it equals 97 per cent for the empirical profile.

The model does match well to the change in 
$q$ seen a larger radii, from  $q\simeq 0$ at $r \la r_{\rm c}$ to $q \simeq 3$ at $r \ga r_{\rm c}$. This is a simple geometrical effect, reproduced also in flat space-time, however, it is of minor importance for the total reflection spectrum as so little of the total reflection signal is produced at these radii. 

The large difference in emissivity from the central regions translates into a clear difference in predicted spectra, as shown in Figure \ref{radial}(b). The extended corona of WF12 does not give the same reflected spectrum as derived from the empirical emissivity profile fit to the data from 1H0707--495.

\section{Spectral analysis}

In Sections \ref{sect:bpl} and \ref{sect:refl} we address W14, therefore, we follow their assumption that reflection  is produced by weakly ionised disc and (still following W14) we adopt parameters of the reflection model (ionisation parameter, $\xi$, iron abundance, Z$_{\rm Fe}$, and $i$) of \citet{2010MNRAS.401.2419Z}.
In Section \ref{sect:absrefl} we relax these assumptions and allow a free fit of our extended corona model to the data. 

\subsection{Phenomenological radial emissivity}
\label{sect:bpl}

Following W14 we fitted the L, M and H spectra assuming a twice-broken power-law radial emissivity, for which we adopted also their constraints on the emissivity parameters (see the footnote of Table \ref{tab:bpl}), and we fixed $\xi=53.4$, $i=53.96 \degr$, Z$_{\rm Fe}=8.88$ and $r_{\rm in}=r_{\rm ISCO}$.
We applied model \texttt{reflkerr} \citep{2019MNRAS.485.2942N}, which computes the reflection spectrum for an arbitrary twice broken power law radial emissivity.
Our results, see Table \ref{tab:bpl}, are consistent with  those of W14. In particular, we find $r_{\rm br,2}$ between  $\simeq 20$ and $\simeq 30$, increasing with  the observed flux. 
The fit quality with spectral binning (2) is also consistent with W14 and seemingly excellent. For binning (1) we found only insignificant differences in the values of the fitted parameters, however, here they give a much worse $\chi^2/{\rm DoF} > 2$. 

We now fit our extended disc corona reflection model directly to the data to explicitly test if this geometry can indeed produce the observed emissivity. 
\texttt{reflkerr} computes the transfer of both the reflected and the direct radiation from the corona, therefore, properly calculates the reflection strength for this geometry. 
We find that the data can only be fit with an unphysically strong reflection component, 
with $\mathcal{R} > 10$. This rules out the parameters of W14 for the 
disc-corona model in 1H0707--495. The empirical twice broken power law 
emissivity is only an approximation of models with a radially extended X-ray corona over the disc, and is quite a poor approximation for models with very steep inner emissivity as these produce much more inner disc reflection than is expected from an isotropic (in its rest frame) radially extended source.

\begin{figure}
\centering
\includegraphics[height=22cm]{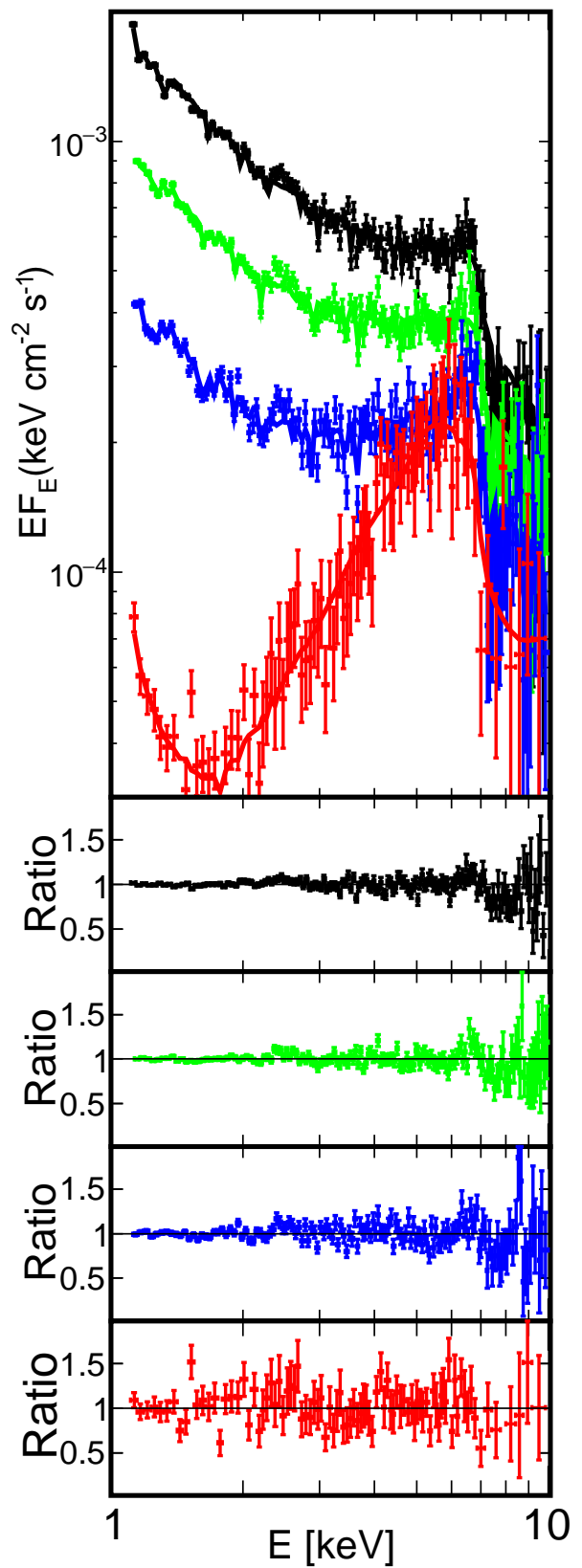}
\caption{The upper panel shows the unfolded data and model spectra for our best-fitted models for VL (red), L (blue), M (green) and H (black) using \texttt{zxipcf*reflkerr\_elp}. 
Spectral binning (1) was used here.
The model parameters are given in Table \ref{tab:x}. 
The model is shown by the solid curve.
The lower panel shows the fit residuals given as the data-to-model ratio for VL, L, M, and H from bottom to top.
}
\label{fiteeuf}
\end{figure}

\begin{figure*}
\centering
\includegraphics[height=10cm]{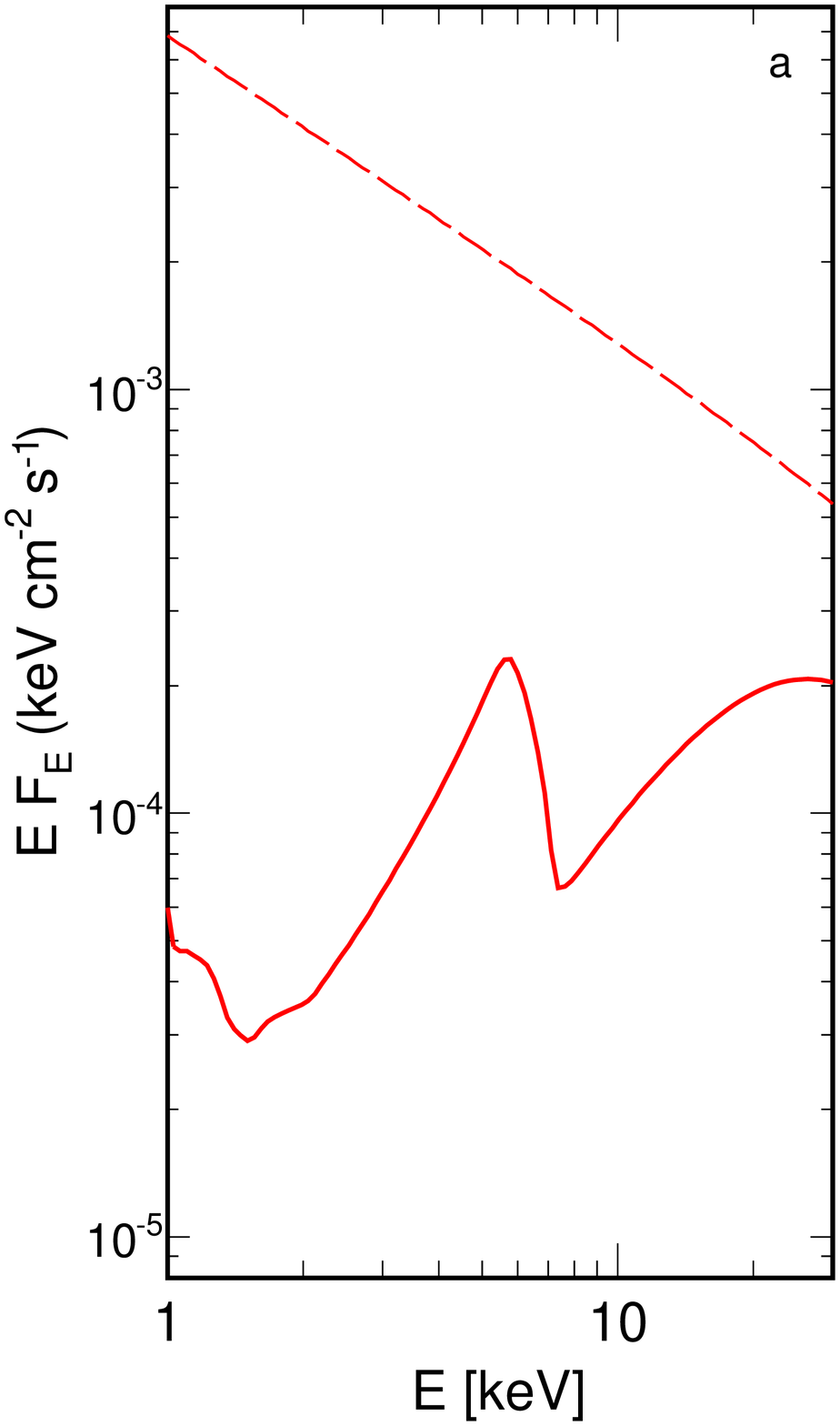}
\includegraphics[height=10cm]{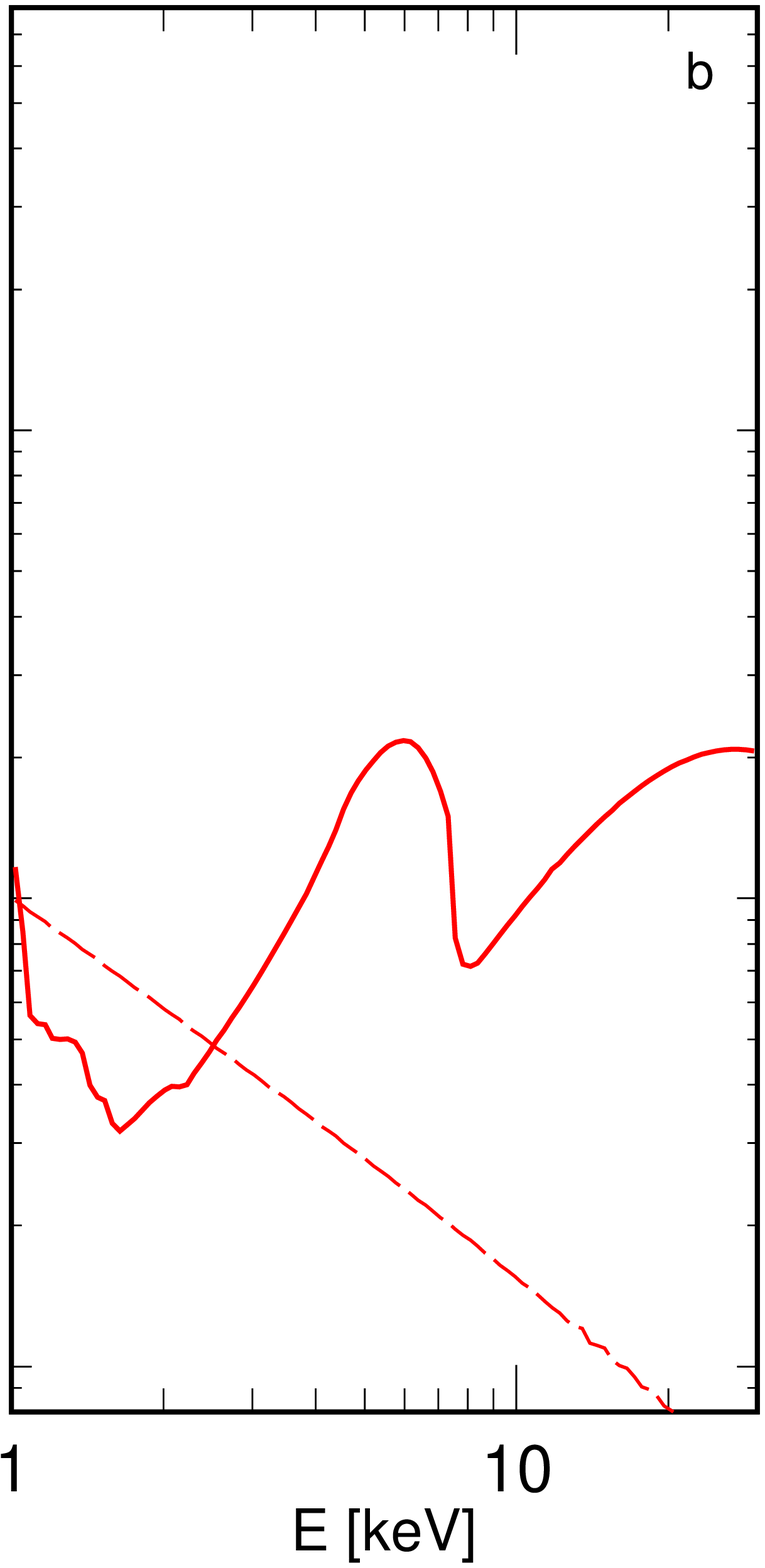}
\includegraphics[height=10cm]{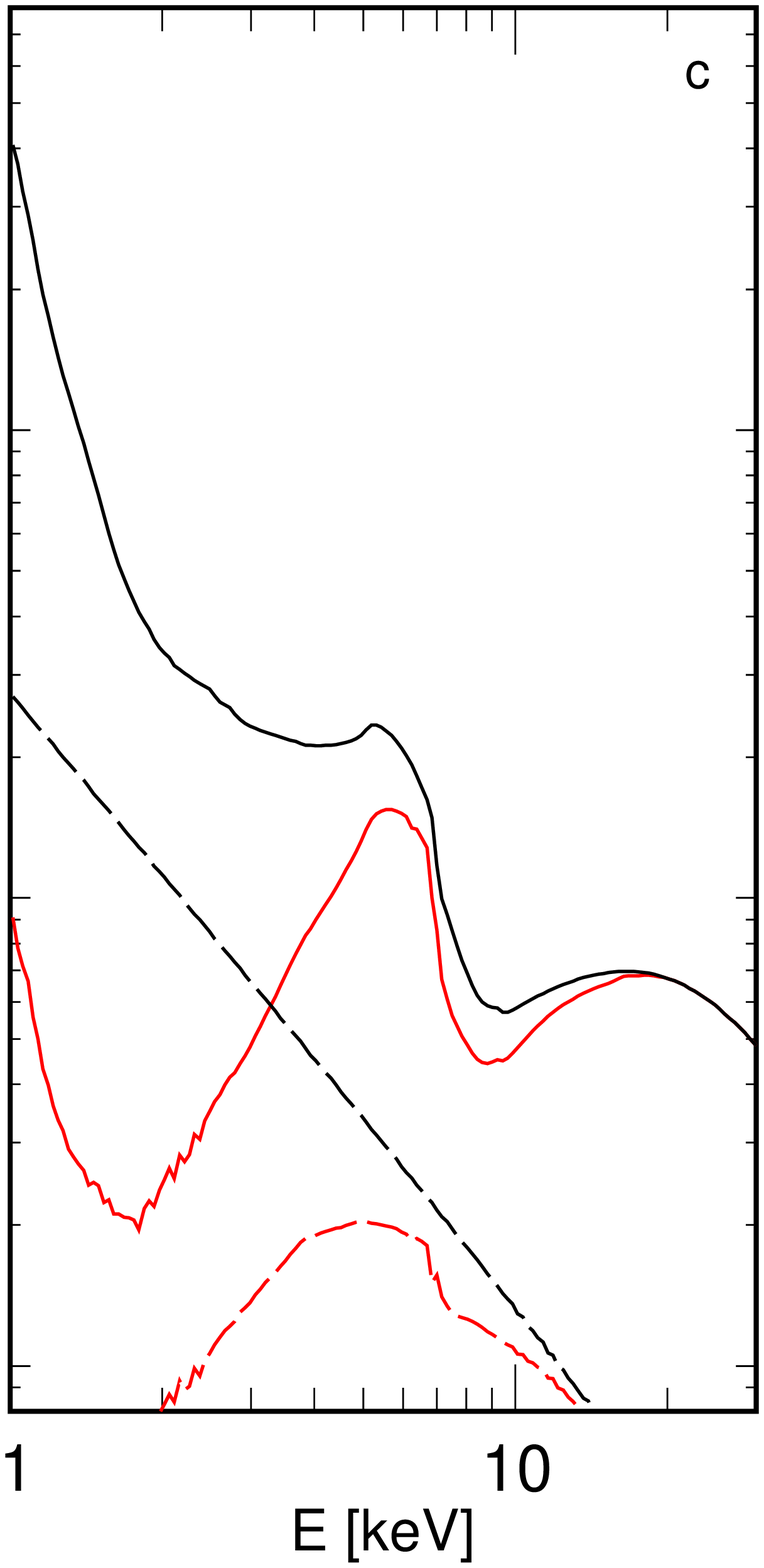}
\caption{The primary (dashed) and reflected (solid) spectra in our best models for VL with (a) \texttt{reflkerr} (Section \ref{sect:bpl}), (b) \texttt{reflkerr\_elp} with a weakly ionised disc (Section \ref{sect:refl}) and (c) \texttt{reflkerr\_elp} with a strongly ionised disc and an ionised absorber (Section \ref{sect:absrefl}). The spectra correspond to the fitted parameters for VL given in Tables \ref{tab:bpl}, \ref{tab:refl} and \ref{tab:x}, respectively, except for $\mathcal{R}$, which here was set to the physical value, i.e.\ $\mathcal{R}=1$. As can be seen in Figure \ref{fiteeuf}, the observed VL spectrum has a strong depression at $\sim 2$ keV. For the models shown in panels (a) and (b), the contribution of the primary component must be artificially reduced, by a factor $>750$ and $> 25$, respectively, to obtain a similar spectral depression in the total spectra. In panel (c) the black curves show the unabsorbed and the red curves show the absorbed spectra. In this case, the relativistic effects combined with absorption  allow to reproduce the observed shape with $\mathcal{R}=1$. 
}
\label{fig:VL}
\end{figure*}

For the VL spectrum, the twice-broken radial profile, with constraints on ($q_1$, $q_2$, $q_3$) of W14, does not give a good fit. We therefore fitted a single-broken profile, see Table \ref{tab:bpl}. The fitted profile is very steep, with $q_2 > 7$ at $r > 2.3$. However, the inner $q_1=-0.8$ indicates a strongly reduced contribution from $r \la 2$. This break in the emissivity is highly significant. Using a single power-law (with $r_{\rm in} = r_{\rm ISCO}$) we get a much worse fit with $\Delta \chi^2 \simeq +50$ for binning (1). The fit also significantly improves after allowing $\xi$ to vary; it then decreases to the minimum value allowed in \texttt{reflionx}, i.e.\ $\xi=10$.
The spectral feature observed in VL between $\sim 2$ keV and $\sim 7$ keV (see Figure \ref{fiteeuf})  can be reproduced by a relativistic line from a weakly ionised disc. However, in the disc-corona model such a spectral description requires an unrealistic reduction of the primary continuum, with  $\mathcal{R} > 750$, see Figure \ref{fig:VL}(a), which again rules out this geometry.

\begin{figure*}
\centering
 \includegraphics[trim={0cm 4.5cm 0cm 4.5cm},clip,height=8.5cm]{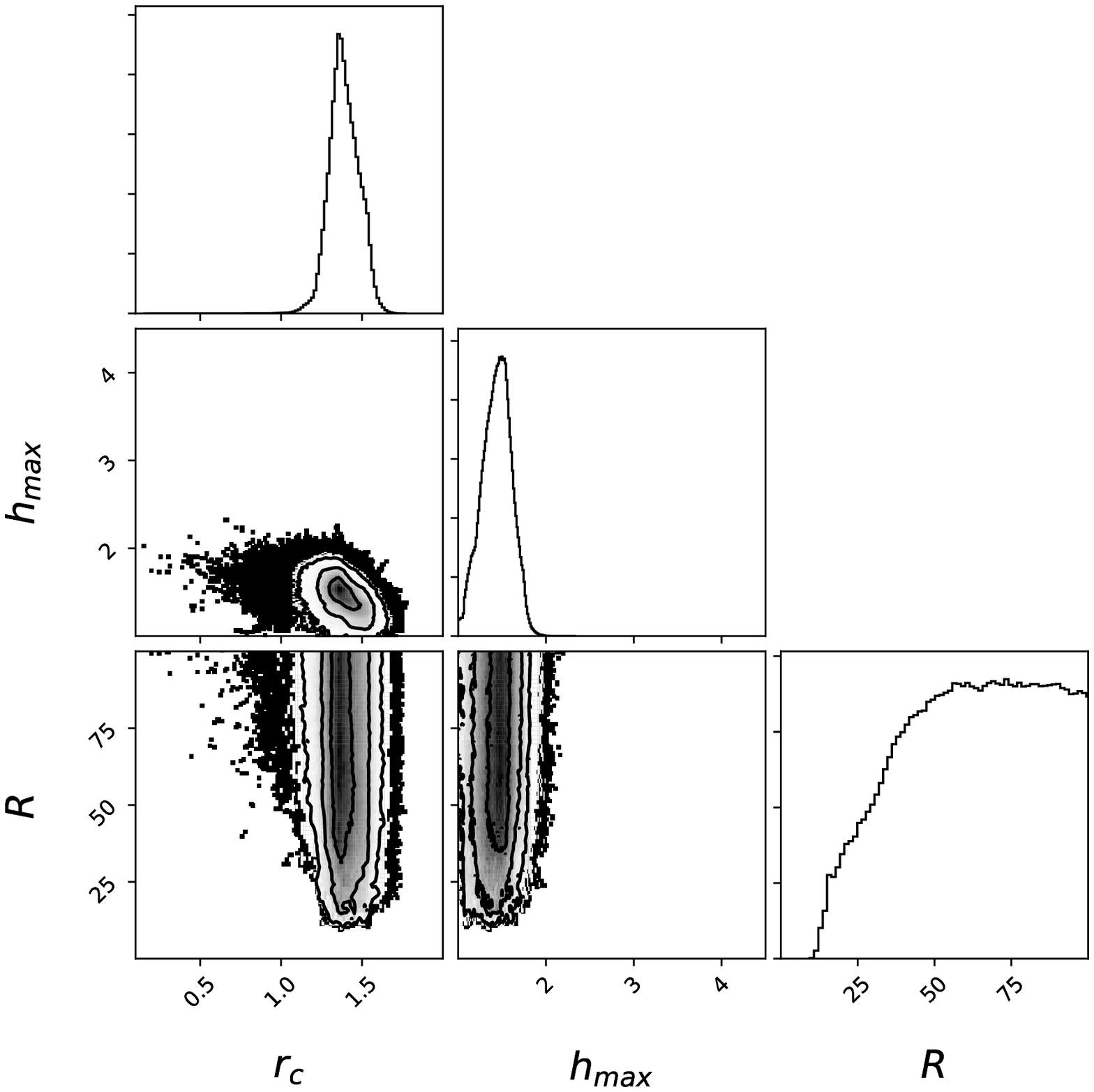}
 \includegraphics[trim={0cm 4.5cm 0cm 4.5cm},clip,height=8.5cm]{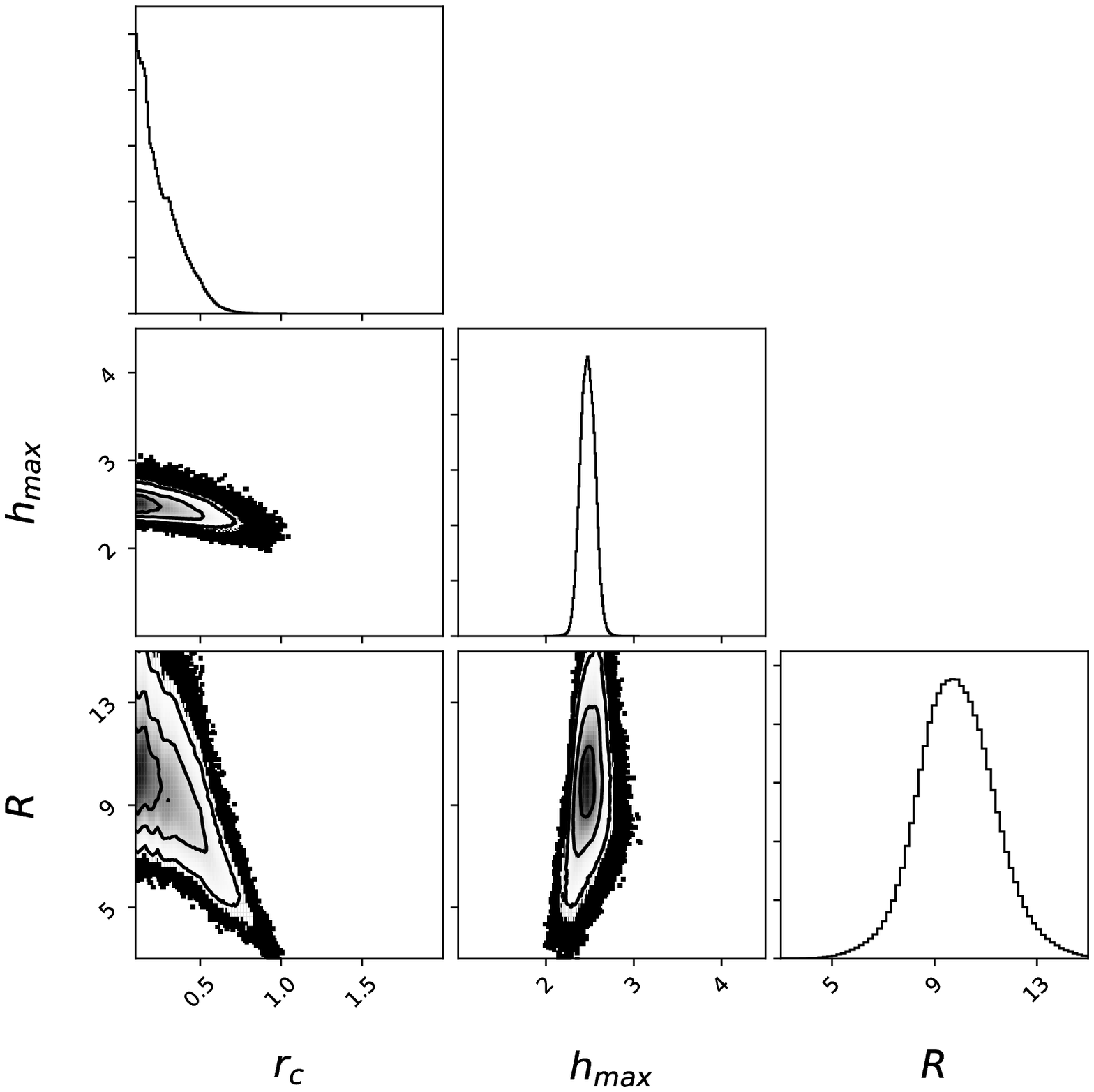}\\
 \includegraphics[trim={0cm 4.5cm 0cm 4.5cm},clip,height=8.5cm]{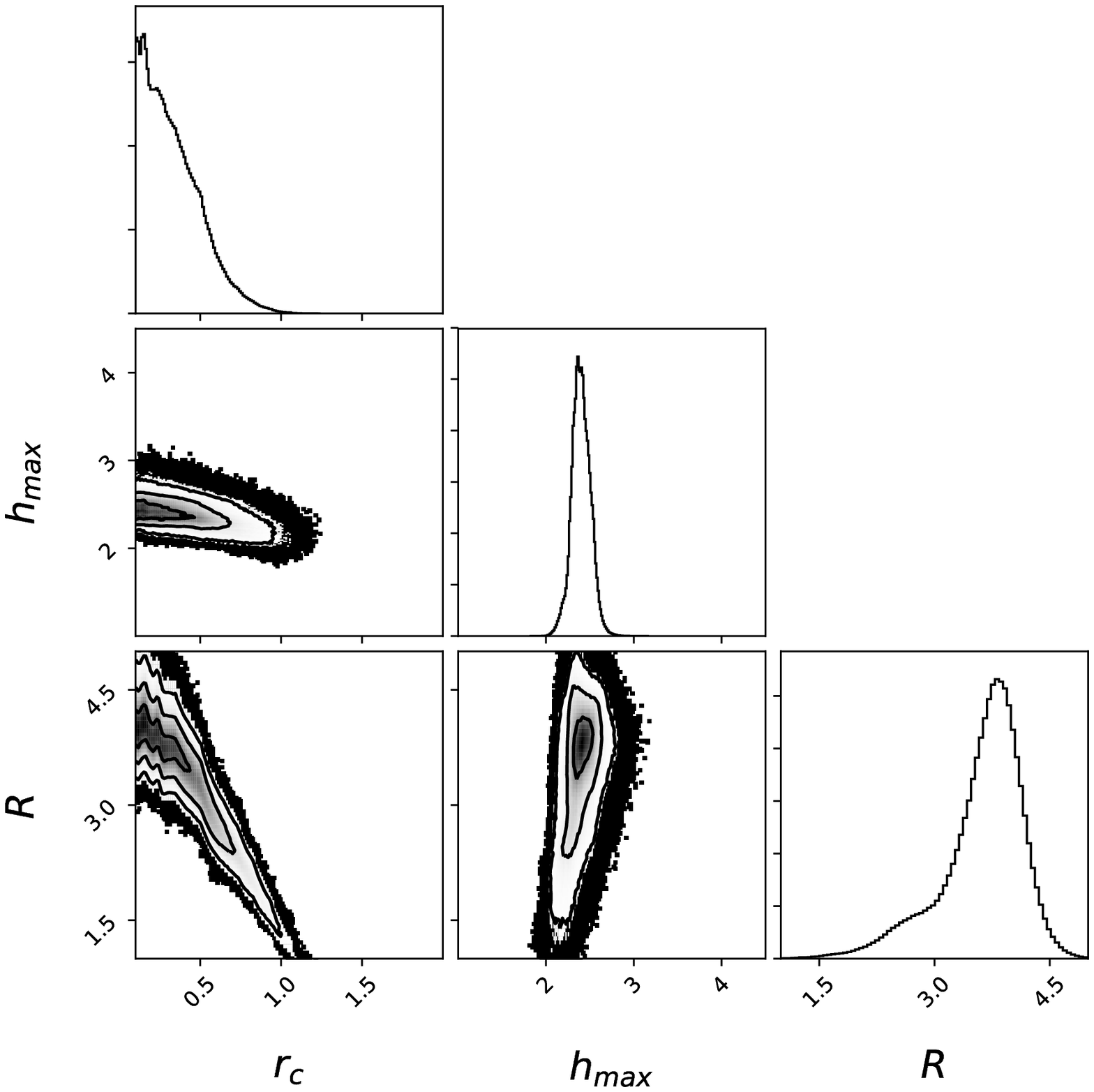}
 \includegraphics[trim={0cm 4.5cm 0cm 4.5cm},clip,height=8.5cm]{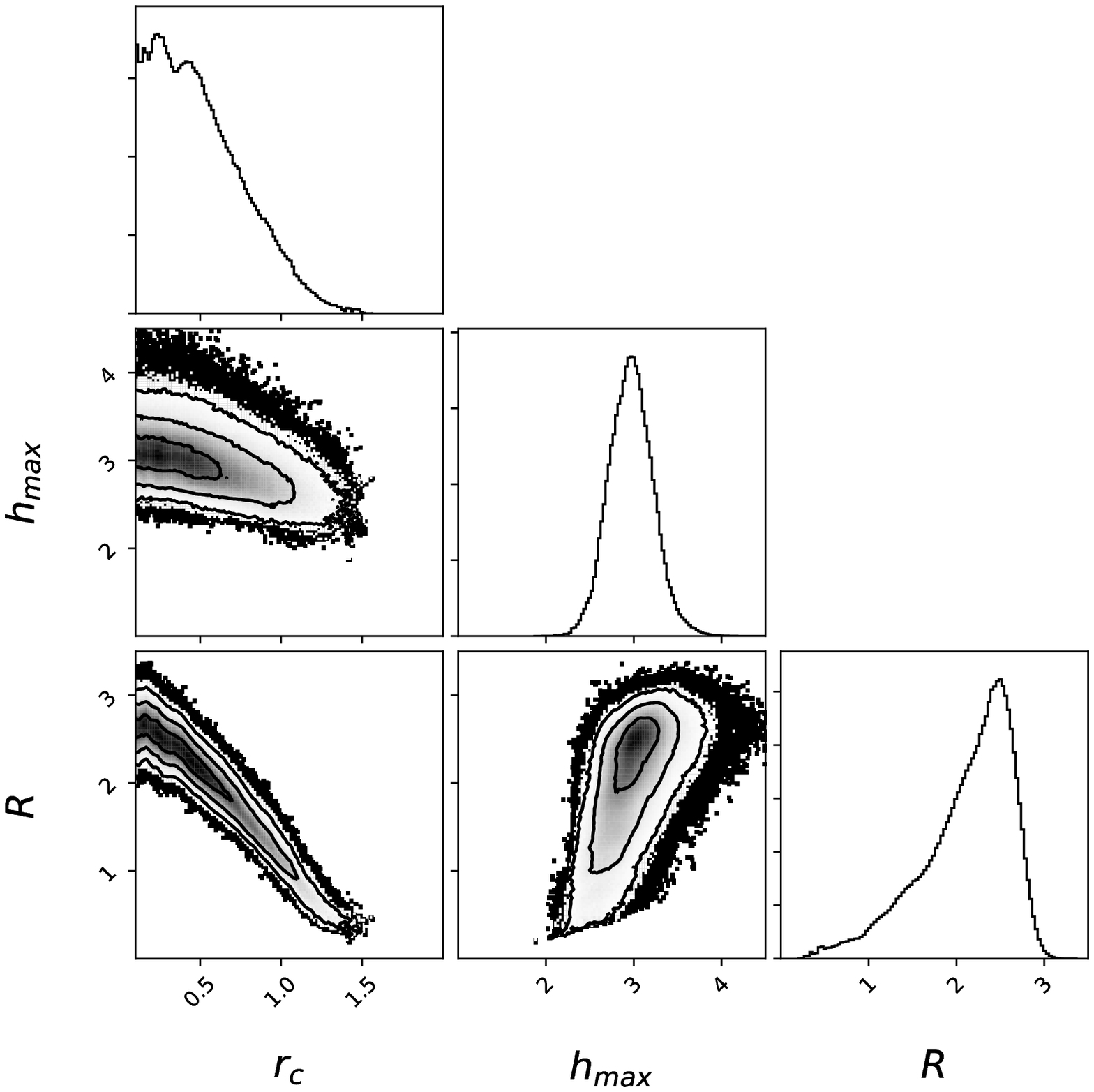}
\caption{Probability distributions showing correlations between the $r_{\rm c}$, $h_{\rm max}$ and $\mathcal{R}$ parameters of the \texttt{reflkerr\_elp} model with a weakly ionised disc (see Table \ref{tab:refl}) fitted in the 1.1--10 keV range to the VL (top left), L (top right), M (bottom left) and H (bottom right) spectrum, obtained in the MCMC analysis using \texttt{xspec\_emcee} implemented by Jeremy Sanders. The results of this analysis are presented using package \texttt{corner} \citep{corner}. The contours in the 2D plots correspond to the significance of $\sigma = 1, 2, 3$. The histograms show the probability distributions for the individual parameters. 
}
\label{fig:mcmc}
\end{figure*}

\begin{table}
\centering
\caption{
The results of spectral fitting of the model \texttt{reflkerr\_elp} with $r_{\rm c}$ set to $r_{\rm br,2}$ given in Table \ref{tab:bpl}. In all models $a=0.998$, $r_{\rm out}=1000$,  $i = 53.96^{\circ}$, $\xi = 53.44$, $Z_{\rm Fe} = 8.88$, $r_{\rm in} = r_{\rm ISCO}$ and $h_{\rm min} = 0$. The best-fit values and parameter uncertainties given in the table correspond to binning (1); the differences in the best-fit values with binning (2) are insignificant.}
\begin{tabular}{ lllll  }
 \multicolumn{4}{c}{Model: \texttt{reflkerr\_elp}} \\
 \hline
                        & L                     & M                     & H     
        \\
 \hline
 $r_{\rm c}$            & $18.5^{(f)}$          & $28.2^{(f)}$          & $32.0^{(f)}$  \\[0.1cm]
 $h_{\rm max}$          & $35.5^{+10.2}_{-9.4}$ & $26.1^{+8.4}_{-7.5}$  & $12.4^{+6.5}_{-5.3}$  \\[0.1cm]
 $\Gamma$               & $2.52^{+0.02}_{-0.05}$& $2.69^{+0.02}_{-0.01}$& $2.84^{+0.01}_{-0.01}$\\[0.1cm]
 $\mathcal{R}$          & $8.1^{+0.9}_{-0.6}$   & $5.5^{+0.4}_{-0.4}$   & $4.8^{+0.3}_{-0.5}$   \\[0.1cm]
 \hline
 & \multicolumn{3}{c}{$\chi^{2}/$DoF} \\
 \hline
 (1)                    & 460/136               & 479/145               & 451/144\\
 (2)                    & 1057/692              & 1329/936              & 1348/954\\
 \hline 
\end{tabular}\\
\label{tab:rbr2}
\end{table}

\subsection{\texttt{reflkerr\_elp} plain reflection}
\label{sect:refl}

We now use our model \texttt{reflkerr\_elp}. We first aim to quantitatively test 
whether irradiation by an 
extended disc corona out to $r_{\rm c}$ can give a broken power law emissivity with 
$r_{\rm c} \simeq r_{\rm br,2}$ as proposed by WF12 and W14. Therefore, (1) we keep here the same assumptions about the rest-frame reflection as in the previous section, 
and (2) we
set $r_{\rm c}$ equal to $r_{\rm br,2}$ of the \texttt{reflkerr} model fitted to a given state (Table \ref{tab:bpl}). This gives much larger values of $\chi^2$ than obtained with \texttt{reflkerr} in Section \ref{sect:bpl}, with $\Delta \chi^2 \simeq (150 - 200)$ for spectra L, M and H with binning (1) as shown in 
Table \ref{tab:rbr2}. There is no good fit for VL as discussed above, so we do not tabulate results for this spectrum. 

The worsening of the fit shows that the 
disagreement between the empirical and  physical emissivity profiles discussed in Section \ref{sect:comp} is important even for the analysis limited to the 1--10 keV range.
Using $r_{\rm br,2}$ to estimate $r_c$ (as proposed by WF12 and W14) does not work for an extended corona geometry. The very large inner emissivities inferred from the twice broken power law require a more radially concentrated X-ray illumination pattern than produced from a uniform corona, even in full General Relativity. 

This is the case even when we allow $r_{\rm c}$ to vary, rather than fixing it to $r_{\rm br,2}$. The inferred reflection in the data is dominated by the inner, extremely steep emissivity so the fitted $r_{\rm c}$ as well as $h_{\rm max}$ are very small for all spectra, as detailed in  Table \ref{tab:refl}. 
This means that the source geometry is now close to that of the standard lamp-post, except that it is rotating. 

For such small $r_{\rm c}$ and $h_{\rm max}$, any value of $h_{\rm min}$ between $r_{\rm hor}$ and $h_{\rm max}$ gives identical results, because 
radiation from $h$ close to $h_{\rm max}$ dominates both the observed and the irradiating flux.
For simplicity, for all results presented here, we assumed $h_{\rm min} = r_{\rm hor}$ for L, M and H. For VL, where the fitted $r_{\rm c}$ is slightly larger, we assumed  $h_{\rm min} = 0$; allowing $h_{\rm min}$ to vary does not change the results.

\begin{table}
\centering
\caption{
The results of spectral fitting of our model \texttt{reflkerr\_elp}. In all models $a=0.998$, $r_{\rm out}=1000$,  $i = 53.96^{\circ}$ and $Z_{\rm Fe} = 8.88$; $h_{\rm min} = 0$ in VL and $h_{\rm min} = r_{\rm hor}$ in L, M and H. The best-fit values and parameter uncertainties given in the table correspond to binning (1); the differences in the best-fit values with binning (2) are insignificant.}
\begin{tabular}{ lllll  }
 \multicolumn{5}{c}{Model: \texttt{reflkerr\_elp}} \\
 \hline
			& VL 		& L 			& M 		& H \\
 \hline
 $r_{\rm c}^{(a)}$  		& $1.5^{+0.1}_{-0.1}$ 	& $0.1^{+0.3}_{-0}$	& $0.1^{+0.5}_{-0}$	& $0.3^{+0.6}_{-0.2}$	\\[0.1cm]
 $h_{\rm max}$		& $1.5^{+0.3}_{-0.2}$	& $2.4^{+0.2}_{-0.3}$	& $2.4^{+0.4}_{-0.4}$	& $3.0^{+0.8}_{-0.6}$	\\[0.1cm]
 $\Gamma$		& $2.77^{+0.04}_{-0.15}$& 3.09$^{+0.07}_{-0.10}$& 3.01$^{+0.06}_{-0.03}$& 3.00$^{+0.02}_{-0.03}$\\[0.1cm]
 $\xi$			& $10^{+2.6}_{-0}$ 	& $53.44^{f}$ 		& $53.44^{f}$ 		& $53.44^{f}$ 		\\[0.1cm]
 $\mathcal{R}$ 		& $> 25$ 		& $10.7^{+2.2}_{-2.4}$	& $3.9^{+0.5}_{-1.2}$ 	& 2.5$^{+0.4}_{-1.7}$	\\[0.1cm]
 $r_{\rm in}$ 		& $1.7^{+0.2}_{-0.1}$  		& $r^{(f)}_{\rm ISCO}$	& $r^{(f)}_{\rm ISCO}$ 	& $r^{(f)}_{\rm ISCO}$	\\[0.1cm]
 \hline
 & \multicolumn{4}{c}{$\chi^{2}/$DoF} \\
 \hline
 (1) 	& 119/86  	& 314/135 		& 313/144 		& 353/143\\
 (2) 	& 183/128  	& 859/691 		& 1113/935 		& 1129/953\\
 \hline 
\end{tabular}\\
\tablefoot{{\it (a)} the minimum value of $r_{\rm c}$ allowed in the \texttt{reflkerr\_elp} model is 0.1.}
\label{tab:refl}
\end{table}

Figure \ref{fig:mcmc} shows the probability distributions and correlations between $r_{\rm c}$, $h_{\rm max}$ and $\mathcal{R}$, obtained using the Markov Chain Monte Carlo (MCMC) method. The major degeneracy concerns the anticorrelation between $r_{\rm c}$ and $\mathcal{R}$ in L, M and H, which is related to the increase of $V$ with $r_{\rm c}$ (for the assumed velocity field close to the symmetry axis, see Section \ref{sect:model}). This is due to 
the increase of reflection strength with $V$, see Figure \ref{fig:v}. 
The VL spectrum requires a negligible contribution of the primary component, with $\mathcal{R} \ga 25$, see Figure \ref{fig:VL}(b), so
for this state we do not get the $\mathcal{R}$-$r_{\rm c}$ anticorrelation.

For VL we allowed also $r_{\rm in}$ to vary and we find that
$r_{\rm in} > r_{\rm ISCO}$ is formally significant; setting  $r_{\rm in} = r_{\rm ISCO}$ gives, for binning (1), $\Delta \chi^2 = 17$.

All these fits are somewhat worse than those of the 
phenomenological broken power law emissivity reflection fits, typically by $\Delta\chi^2\sim 10-20$ for binning (1). This is because the model geometry is now reproducing the inner, very steep emissivity (where most of the reflection signal is produced) at the expense of the weak outer reflection component. More concerning is that 
they all require  $\mathcal{R}>1$ so are physically unrealistic except for the brightest state (H). 

The pure reflection models clearly show that the
best fit extended source geometry is not very extended for all the spectra of 1H0707--495. The illumination pattern derived from these fits requires a very centrally concentrated X-ray corona, close to the lamp-post geometry except for its rotation kinematics modifying the illumination pattern. Hence there is also no evidence for the size of the corona being correlated with the luminosity, as claimed in W14.

\begin{table}
\centering
\caption{
The results of spectral fitting of the model \texttt{reflkerr\_elp}, with a partially covering ionised absorption computed with \texttt{zxipcf}. Negative values of the redshift parameter in \texttt{zxipcf} give the blueshift corresponding to an outflow of the absorber. In all models $a=0.998$ and $r_{\rm out}=1000$; $h_{\rm min} = 0$ in VL and $h_{\rm min} = r_{\rm hor}$ in L, M and H. The model is fitted jointly to the four states with linked $i$ and $Z_{\rm Fe}$. We give $\chi^{2}/$DoF for the joint fit and contribution of individual states to the total $\chi^{2}$.
The best-fit values and parameter uncertainties given in the table correspond to binning (1); the differences in the best-fit values with binning (2) are negligible.}
\begin{tabular}{lllll}
\multicolumn{5}{c}{Model: \texttt{zxipcf*reflkerr\_elp}} \\[0.1cm]
\hline
             & VL            & L                     & M             & H \\[0.1cm]
\hline
 & \multicolumn{4}{c}{\texttt{zxipcf}}\\
\hline
 $n_H$ & $17^{+2}_{-2}$ & $113^{+69}_{-73}$ & $51^{+42}_{-36}$ & $23^{+31}_{-14}$   \\[0.1cm]
 $\log_{10}(\xi) $ & $1.6^{+0.9}_{-0.3}$& $3.9^{+0.2}_{-0.4}$& $4.0^{+0.1}_{-0.2}$ & $3.9^{+0.1}_{-0.4}$ \\[0.1cm]
 $f_{\rm cov}$ & $0.98^{+0.01}_{-0.01}$ & $1.0^{+0}_{-0.13}$& $1.0^{+0}_{-0.16}$& $1.0^{+0}_{-0.45}$  \\[0.1cm]
 $z$ & $0.04^{+0}_{-0.03}$ & $-0.09^{+0.01}_{-0.01}$& $-0.10^{+0.01}_{-0.01}$ & $-0.12^{+0.01}_{-0.01}$ \\[0.1cm]
\hline
 & \multicolumn{4}{c}{\texttt{reflkerr\_elp}}\\
\hline
 $r_{\rm c}$  & $1.7^{+0.5}_{-0.4}$  & $1.1^{+0.1}_{-0.1}$ & $0.6^{+0.1}_{-0.2}$ & $0.7^{+2.2}_{-0.6}$   \\[0.1cm]
 $h_{\rm max}$ & $1.8^{+1.1}_{-1.2}$   & $3.8^{+1.3}_{-0.4}$  & $4.0^{+0.5}_{-0.6}$   & $4.8^{+1.0}_{-0.9}$   \\[0.1cm]
 $i \; [^{\circ}]$ & \multicolumn{4}{c}{ $39.7^{+0.9}_{-1.3}$} \\[0.1cm]
 $Z_{\rm Fe}$      & \multicolumn{4}{c}{ $5.0^{+0.4}_{-0.3}$} \\[0.1cm]
 $\Gamma$ & $3.27^{+0.03}_{-0.17}$ & $2.38^{+0.02}_{-0.02}$ & $2.55^{+0.01}_{-0.01}$ & $2.68^{+0.01}_{-0.01}$\\[0.1cm]
 $\xi$    & $5010^{+2140}_{-2690}$ & $1050^{+110}_{-60}$ & $2120^{+180}_{-300}$ &  $3850^{+920}_{-510}$ \\[0.1cm]
 $\mathcal{R}$ & $>0.6$ & $1.4^{+1.0}_{-0.3}$  & $1.5^{+1.2}_{-0.1}$   & $1.0^{+0.1}_{-0.2}$   \\[0.1cm]
 $r_{\rm in}$  & $2.6^{-0.5}_{-0.5}$    & $1.4^{+0.5}_{-0.2}$ & $1.4^{+0.2}_{-0.2}$   & $1.3^{+0.2}_{-0.1}$    \\[0.1cm]
 \hline
 & \multicolumn{4}{c}{$\chi^{2}/$DoF} \\[0.1cm]
 (1)                    & \multicolumn{4}{c}{671/484} \\[0.1cm]
      & 92        & 202               & 182               & 196\\[0.1cm]
 \hline
 (2)                    & \multicolumn{4}{c}{2841/2682} \\[0.1cm]
      & 146      & 742               & 977              & 976\\[0.1cm]
 \hline
\end{tabular}
\tablefoot{$n_H$ is given in the unit of $10^{22}$ cm$^{-2}$.}
\label{tab:x}
\end{table}

\subsection{\texttt{reflkerr\_elp} reflection + ionised absorption}
\label{sect:absrefl}

We now include photoionised ionised absorption in addition to our model for reflection from the extended disc-corona, 
\texttt{reflkerr\_elp}, as multiple studies have shown that this is significantly present \citep{2012MNRAS.422.1914D,2016MNRAS.461.3954H,2018MNRAS.481..947K}. We first model this absorption using the 
\texttt{zxipcf} model implemented in \texttt{xspec}, though we note that this assumes only a rather small turbulent velocity broadening of 200 km/s, much smaller than those inferred from the data (around 10,0000 km/s in \citealt{2018MNRAS.481..947K}). It 
also assumes solar abundances for all elements, whereas the reflection fits typically require large overabundance of iron, and uses ion population calculated from a spectrum which has $\Gamma=2.2$, rather harder than that inferred for 1H0707--495. 
Nonetheless, it gives a simple model to approximately characterize some of the spectral complexity associated with partially ionised absorption. 

We do not impose any constraints on the model parameters except for linking the reflection 
$i$ and Z$_{\rm Fe}$  across all four states, and fixing $a=0.998$.
The spectral fitting results are given in Table \ref{tab:x}, Figure \ref{fiteeuf} shows the the spectra and best-fit models to all states. 

Adding the photoionised absorber significantly improves the spectral description in all four states,
both compared to the extended disc-corona fits with free corona size shown above, and even compared to the phenomenological twice broken power law emissivity fits. 
We allow here the scaling parameter of reflection, $\mathcal{R}$, to vary but all fits correspond to a physically self-consistent $\mathcal{R} \simeq 1$. This is even the case for the VL spectrum, which always previously required 
$\mathcal{R} \gg 1$. 

Figure \ref{fig:VL} shows the sequence of model fits to the 
VL spectrum, with the direct power law (dashed) and reflection component (solid) shown separately. 
Panel (a) shows the low ionisation \texttt{reflkerr} fit with reflection calculated from the phenomenological emissivity as in Section 5.1, but for a reflection normalisation set to the physical $\mathcal{R} = 1$ rather than the extreme values required from the fits. Even though the emissivity best fit is a single, very centrally concentrated power law, gravitational effects alone are not sufficient to enhance the reflection to the extent required by the data. 
Panel (b) shows the fit with reflection calculated from the self consistent illumination pattern of an extended, rotating corona. The disc ionisation parameter is similarly low to that derived from the phenomenological emissivity fits, so the reflected component shape is much the same, but its amplitude is now dramatically enhanced by the source kinematics, with the intrinsic rotation changing the illumination pattern on the disc. However, this is still insufficient to explain the inflection in the VL data at 2~keV. Instead, Panel (c) 
shows the fit including ionised absorption. Unlike the L, M and H fits, the ionised absorber in VL is only moderately rather than highly ionised ($\xi\sim 40$ rather than $10^4$)
Its effect on the spectrum is to produce strong continuum curvature rather than distinct ionised absorption line features, with a strong low energy cutoff below 4~keV from bound-free transitions in metals such as Si, S and Fe L. This continuum curvature gives a much better fit to the data, and changes the shape of the best fit reflection spectrum so that it now is better described by a highly ionised disc. This increases the amount of reflection at low energies compared to the Compton hump, as well as shifting the line and edge energies. The total continuum spectrum then has a strong dip at 2~keV as well as at 9~keV, as observed in the data (see also Figure \ref{fiteeuf}). 

Figure \ref{fiteeuf} also shows the other datasets fit with this model including ionised absorption. The change in behaviour of the continuum at low energies between VL and L, M and H is obvious. L, M and H instead give best fit ionised absorption which is highly ionised $\xi\sim 10^4$. 

For all the datasets, the geometry of the extended source is actually very little changed from that derived in the previous section. Both $r_{\rm c}$ and $h_{\rm max}$ are 
small, making the model much more like a lamp-post (but with rotation) than a corona with large scale extent over the disc.
Similarly as in Section \ref{sect:refl}  we assume $h_{\rm min} = r_{\rm hor}$ for L, M and H and $h_{\rm min} = 0$ for VL; allowing $h_{\rm min}$ to vary does not change our results. 

However, despite the lack of change in source geometry,
all the spectral fits are now consistent with 
$\mathcal{R} \simeq 1$. This is 
due to the much larger inferred 
ionisation state of the reflector, with  $\xi > 1000$, in all four states instead of $\xi=10-60$ in fits without the additional ionised absorption. The disc albedo is strongly enhanced by ionisation, which enhances the  
reflection spectrum in soft X-rays for the same radial and vertical extent of the X-ray source. 

The ionised absorber in L, M and H is highly ionised, so 
has little continuum absorption and is instead 
dominated by ionised line features. The inferred columns are extremely high, with $N_h\sim 1.1\times 10^{24}$~cm$^{-2}$ for L, which is optically thick to electron scattering. The inferred columns for M and H are lower, but only by a factor 2-5. Scattering of the continuum from the ionised absorbing material must be important at these columns, so the model is not self consistent. These derived columns could be overestimated as they depend strongly on the 
velocity structure of the absorbing material. Absorption lines saturate when the 
core of the line becomes optically thick i.e. when photons at the line centre cannot escape at the rest energy of the transition. Increasing the column beyond this cannot lead to much more absorption as there are no more photons at the line center to remove. Absorption increases in the wings of the line but Doppler wings are very steep so the line equivalent width does not increase much as the column increases. A larger spread of velocity in the absorbing material gives a wider but shallower line for the same column, so it remains optically thin at higher columns. Fits to 1H0707--495 by \citet{2018MNRAS.481..947K} indicate that the velocity width of the line is around 13,000 km/s as opposed to the 200 km/s used here, which reduces the columns to around $2-4\times 10^{23}$~cm$^{-2}$. However, this still has an optical depth to electron scattering of $\sim 0.2$. The geometry of the material then determines how much the scattered emission contributes to the spectrum, and it is not just scattered continuum but also resonance line scattering. Calculating the contribution of these
cannot be done analytically for non-spherical geometries. Instead it requires a Monte Carlo radiation transfer code. 

\begin{table}
\centering
\caption{
The results of spectral fitting in the 2 -- 10 keV range of the model \texttt{reflkerr}, with wind absorption and scattering computed with \texttt{monaco}. In all models $q=3, a=0.998$ and $r_{\rm out}=1000$. \texttt{monaco} was computed for a fixed wind geometry and we assumed the same inclination angles for the disc and the wind, so this component does not add any free parameters in our fitting procedure. The model is fitted jointly to the three states with linked $i$ and $Z_{\rm Fe}$. We give $\chi^{2}/$DoF for the joint fit and contribution of individual states to the total $\chi^{2}$. The best-fit values and parameter uncertainties given in the table correspond to binning (1); the differences in the best-fit values with binning (2) are negligible.}
\begin{tabular}{llll}
\hline
 \multicolumn{4}{c}{Model: \texttt{monaco*reflkerr}} \\[0.1cm]
 \hline
      & L  & M & H \\[0.1cm]
 \hline
 $r_{\rm in}^{(a)}$  & $770^{+\infty}_{-485}$ & $718^{+\infty}_{-538}$ & $755^{+\infty}_{-587}$ \\[0.1cm]
 $i\; [^{\circ}]$    & \multicolumn{3}{c}{$61.1^{+0.2}_{-2.7}$} \\[0.1cm]
 $Z_{\rm Fe}$        & \multicolumn{3}{c}{$5.3^{+1.7}_{-0.4}$} \\[0.1cm]
 $\Gamma$      & $2.31^{+0.03}_{-0.03}$ & $2.61^{+0.03}_{-0.03}$ & $2.78^{+0.02}_{-0.03}$ \\[0.1cm]
 $\xi$       & $210^{+20}_{-20}$ & $498^{+7}_{-100}$ & $520^{+50}_{-50}$ \\[0.1cm]
 $\mathcal{R}$ & $7.3^{+1.6}_{-1.3}$ & $6.3^{+0.9}_{-0.8}$ & $6.3^{+0.7}_{-1.2}$ \\[0.1cm]
 \hline

 \multicolumn{4}{c}{$\chi^{2}/$DoF} \\[0.1cm]
 (1) & \multicolumn{3}{c}{$509/344$} \\[0.1cm]
        & 159 & 176  & 174 \\[0.1cm]
 \hline
 (2) & \multicolumn{3}{c}{$2142/2028$} \\[0.1cm]
        & 553 & 765  & 825 \\[0.1cm]
 \hline
\end{tabular}\\
\tablefoot{{\it (a)} We use symbol '$+\infty$' to denote $r_{\rm in}$ approaching $r_{\rm out} = 1000$.}
\label{tab:monaco}
\end{table}

\begin{figure}
  \centering
  \includegraphics[height=10.5cm,angle=-90.0]{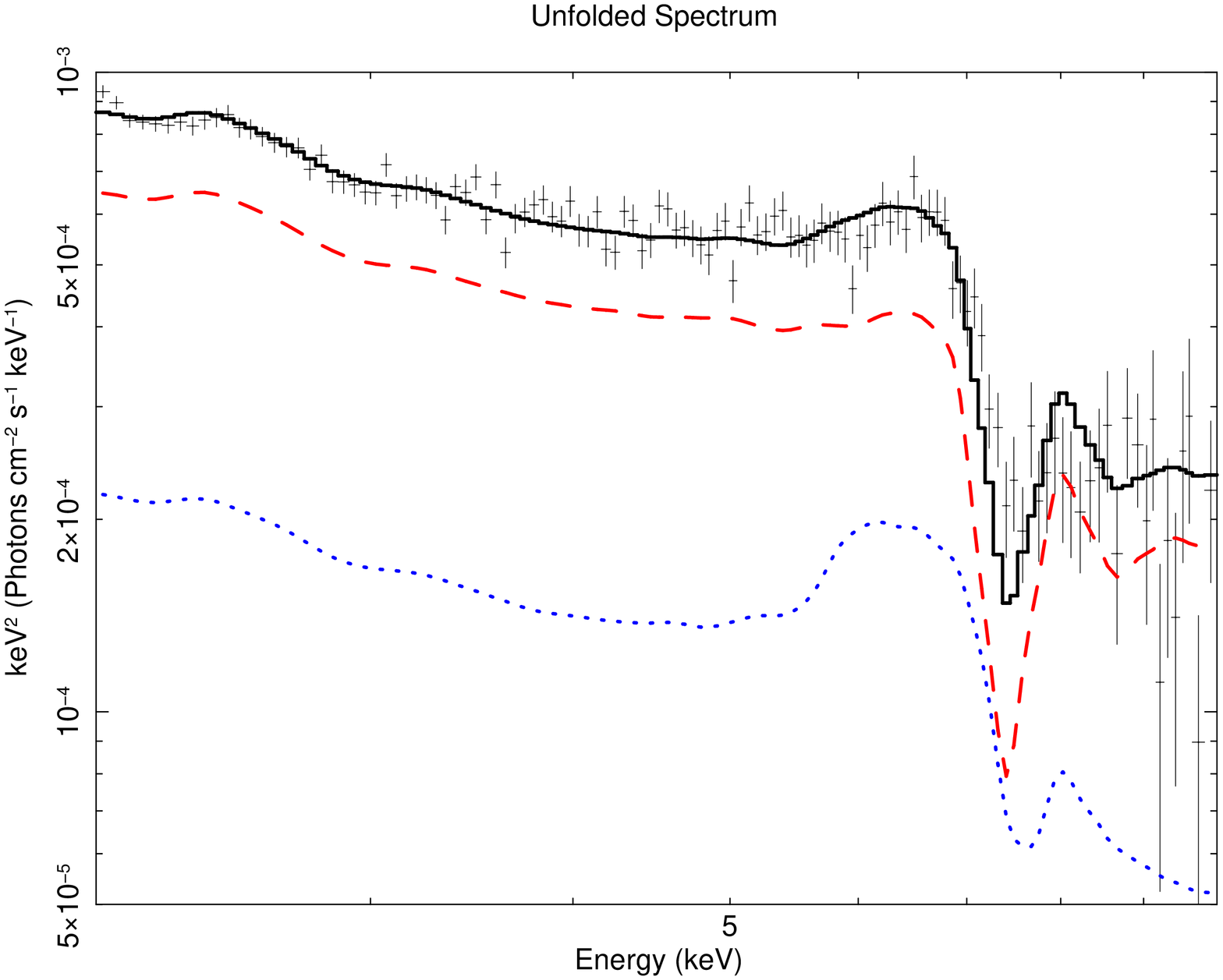}
  \caption{The unfolded data and model spectrum for our best-fitted model for  H  using \texttt{monaco*reflkerr}. The red curve shows the total (i.e.\ primary power-law + reflection) transmitted component attenuated by absorption. The blue curve shows the total scattered component.}
  \label{fig:H_monaco}
\end{figure}

\subsection{\texttt{reflkerr\_elp} reflection + wind absorption and scattering}
\label{sect:wind}

Here we apply the Monte Carlo radiation transport code \texttt{monaco} \citep{2011ApJ...740..103O} to model the absorption and scattering self consistently from a bipolar wind geometry \citep{2015MNRAS.446..663H,2016MNRAS.461.3954H,2017MNRAS.468.1442H,2018MNRAS.476.1776T}. We use the specific bipolar wind simulation of \citet{2019MNRAS.482.5316M}, which is tailored to fit to the time average spectrum of 1H0707--495 (similar to the M and H spectra used here). The model assumes a black hole mass of $2\times 10^6$~M$_\odot$, with a wind filling an   axi-symmetric bicone between $45-56.3^\circ$ (so that the solid angle from the focal point $d$ below the black hole is $\Omega/2\pi = 0.15$), launched at $d = r_{\rm min} = 50$, corresponding to an escape velocity of 0.2c. The mass loss rate in the wind is set to $\dot{M}_{\rm wind}/\dot{M}_{\rm Edd}=0.2$ and this, combined with the velocity law, gives the density structure along the wind streamlines. This is illuminated by a power law with $\Gamma=2.6$ and $L_{\rm X}=0.003L_{\rm Edd}$ to get the self consistent ionisation structure. \citet{2019MNRAS.482.5316M} show that the scattered spectrum in this simulation can also account for the time lags seen in this source around the iron line energy. 

We now use this combined absorption and scattering wind model to modify the intrinsic emission and its reflection from the disc. We model the latter using  \texttt{reflkerr} with a phenomenological emissivity fixed to a single power law with $q=3$. This is justified as after adding the wind 
component we find that the inner radius of the reflecting material increases to a very large value, implying only a very mild relativistic distortion of the reflected spectrum. The fits now have to be restricted to energies above 2~keV due to the energy band limitations of the tabulated model. 

We could not get a good fit for the VL spectrum, as we no longer are including the less ionised material (probably clumps within the wind) which produces the very marked spectral curvature in these data. These less ionised clumps are probably still present in L, M and H, but at a lower level (smaller covering fraction, see \citealt{2016MNRAS.461.3954H}) so the spectrum can be fit without including them. \citet{2016MNRAS.461.3954H} show that the brightest, steepest spectra are least affected by these, so we plot the best fit for H in Figure \ref{fig:H_monaco}. All parameters are given in Table \ref{tab:monaco}, but the reduced energy range means that the goodness of fit cannot be directly compared with previous models. Hence we also refit the \texttt{zxipcf*reflkerr\_elp} model to the same energy range; the $\chi^2/{\rm DOF}$ for the joint fit to L, M and H (with linked $i$ and $Z_{\rm Fe}$) is 369/326 for binning (1). Hence the fit with the wind model is somewhat worse in terms of $\Delta\chi^2$, but the number of degrees of freedom in the model are also lower as it is computed only for a single fixed wind geometry.

Including the scattered emission from the wind dramatically changes the requirement for reflection from the disc to be distorted by extreme relativistic effects. However, the model is still not self consistent as the inferred amount of reflection from the disc is strongly enhanced over that expected, with $\mathcal{R} \sim 6$ rather than the $\simeq 1$. Imposing the constraint of $\mathcal{R} = 1$ we find that small values  of both the inner radius of the disc and the source size are again required. The model also has very different assumptions about metallicity, with Z$_{\rm Fe}=5$ fitted for the reflector, compared to solar in the wind. 

\section{Summary}

This paper shows the model of reflection from an extended corona geometry which can be directly fit to X-ray spectral data. All previous models assumed either a point-like source (lamp-post) or some phenomenological radial emissivity. WF12 associated a twice broken power law emissivity with the illumination pattern expected from a uniform extended corona, with the outer break indicating the radial extent of the corona, a uniform illumination interior to this and then a marked enhancement at small radii due to lightbending and other GR effects. 
However, we show that this qualitative match to extended corona models is not a good quantitative fit to real data. We use 1H0707--495 as apparently the most extreme relativistic reflection source, and directly fit it with both a twice broken power law phenomenological emissivity and the full illumination pattern produced by an extended disc-corona. The outer break radius of the emissivity, $r_{\rm br,2}\sim 30$, so according to WF12 indicates a corona of size $r_{\rm c} = 30$. Explicitly fitting a uniform extended corona with $r_{\rm c}=30$ to the data 
produces a reflection spectrum which strongly deviates from that observed for any scale height because the self consistent centrally peaked illumination pattern is not as steep as required by the phenomenological fits.

Instead, allowing the size (as well as the height) of the corona to be a free parameter makes the corona collapse in size. The best fit corona geometry derived from these fits to the data is a very compact source, with size $\la 1 R_{\rm g}$, and located at most $\sim 2 R_{\rm g}$ from the black-hole horizon in all flux states. These fits show a corona which is highly centrally peaked, rather than being uniformly extended.
This is close to the original lamp-post model used initially for computational convenience, though we do also point out the importance of the source rotation in our model, which makes quite a large increase in the amount of relativistic reflection which escapes to infinity. 

The fitted location of the X-ray source is interestingly consistent with that estimated previously by fitting the lamp-post model to the time-averaged \citep{2012MNRAS.422.1914D} and time-dependent \citep{2018MNRAS.480.2650C} spectra in 1H0707--495, which in both cases gives $h \simeq 3$. The latter, however, depends on the black hole mass, and the lamp-post model of \citet{2018MNRAS.480.2650C} cannot explain the measured delay spectra if it is  $5.3 \times 10^6 M_{\odot}$ \citep{2016ApJ...819L..19P} rather than $2  \times 10^6 M_{\odot}$ estimated in earlier works.

In our analysis we considered only a fixed, maximum value of black hole spin, $a=0.998$. Such a high value was estimated in the lamp-post model by \citet{2012MNRAS.422.1914D} and indeed the fitted values of the inner radius of the disc, $(1.3 - 1.4) R_{\rm g}$ (Section \ref{sect:absrefl}), indicate that $a > 0.99$ is required in this version of the model.

We do not find any sizable changes of the corona geometry between different X-ray flux levels, which would lead to varying distortion of the reflection component, such as shown in Figure \ref{size}. The spectral variations appear to be mostly due to the change of both the amplitude and slope of the primary emission spectrum and the change of the disc ionisation state. We note some hints for geometrical changes only for a very low state, with a weak truncation of the disc at about $2 R_{\rm g}$ (which is ruled out at higher states) and a weak decrease of the height and increase of the radius. However, the large difference between this very low and higher flux spectra is primarily due to the difference of the ionisation state of the absorber.

Our fits then strongly favour a compact source on the spin axis of the black hole. Yet it is difficult to produce a physical model of such a source. There are strong limits on the amount of X-rays it could produce by Comptonization of disc photons \citep{2016AN....337..441D}, and the radiative efficiency is strongly reduced due do photon trapping by the black hole \citep{2016ApJ...821L...1N}.

Including the effect of simple models of ionised absorption from a wind \citep[see also][]{2012MNRAS.422.1914D,2018MNRAS.481..947K} does not change the conclusions about the source size fit to the data, but does indicate that these models are not self consistent as the large columns of ionised material inferred will give a substantial contribution to the spectrum via electron scattering and line emission and resonance line scattering. We use the specific bipolar wind Monte-Carlo calculation of Mizumoto et al (2019) and find that the inclusion of the scattered flux from the wind can dramatically change the parameters. With this model, the reflection spectrum from the disc is only mildly distorted by relativistic effects, typical of material down only to $700 R_{\rm g}$ rather than requiring most of the reflection to come from within $5 R_{\rm g}$, though the fit is somewhat worse than for the phenomenological absorption only models.

Nonetheless, even this model is incomplete as the inferred amount of reflection from the disc is much larger than 
expected in this geometry, and the metallicities of the reflector and wind are different, and most importantly, we neglect any additional absorption from clumps of less ionised material entrained in the hot wind. While such winds cannot be easily modelled {\it ab initio}, they may hold the key to the dramatic spectral features seen in the most extreme AGN. 

\begin{acknowledgements}
We thank Andrzej Zdziarski and the referee, Thomas Dauser, for valuable comments. This research has been supported in part by the Polish National Science Centre grants 2014/13/B/ST9/00570, 2015/18/A/ST9/00746 and 2016/21/B/ST9/02388.  A.N.\ is a member of the International Team 458 at the International Space Science Institute (ISSI), Bern, Switzerland, and thanks ISSI for the support during the meeting in Bern.  M.M.\ acknowledges support from Japan Society for the Promotion of Science (JSPS) overseas research fellowship and Hakubi project at Kyoto University.
\end{acknowledgements}

\bibliographystyle{aa}
\bibliography{1h0707_aa}{}

\begin{appendix}
\section{\texttt{reflkerr\_elp}}
\label{app:a}

\begin{figure}
  \centering
  \includegraphics[height=5.5cm]{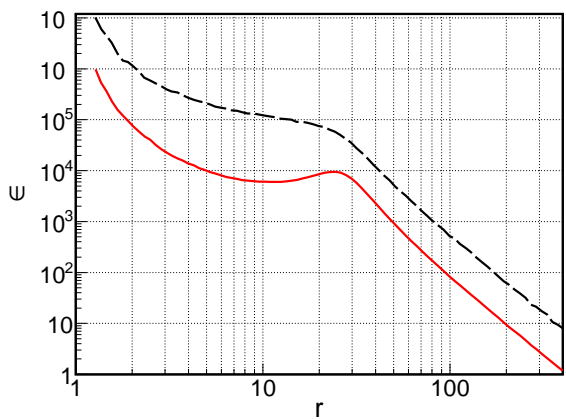}
  \caption{Radial emissivity profiles for a point-like and an extended source, to be compared with Figure 10 in WF12.
The dashed black curve is for a disc-like source with $h=10$ and $\rho \le 25$. 
The red solid curve is for a point-like source at $h=10$ and $\rho = 25$.}
  \label{fig:wf10}
\end{figure}

The construction of the \texttt{reflkerr\_elp} model\footnote{The model can be downloaded at \url{https://users.camk.edu.pl/mitsza/reflkerr_elp}} closely follows that of \texttt{reflkerr\_lp}. In particular, the convolution
of the general-relativistic effects with the rest-frame radiation spectra makes use of the transfer functions, which, following
\citet{1990MNRAS.242..560L} and \citet{1991ApJ...376...90L}, are constructed by tabulating a large number of photon trajectories. The transfer functions are computed only for a single value of the spin parameter, $a=0.998$. The source-to-disc and source-to-observer transfer functions are tabulated in a grid including 51 bins in both $\rho$ and $h$. We use 50 bins logarithmically spaced between $\rho=0.1$ and 50, and the additional bin for $\rho = [0, 0.1]$, and similarly 50 bins logarithmically spaced between $h=0.1$ and 50, and the additional bin for $h=[0,0.1]$. Then, in our parametrization of the model, both the $r_{\rm c}$ and $h_{\rm max}$ parameters can be fitted in the range of $[0.1, 50]$. 
To compute these transfer functions, we generated photon trajectories sampling the region at $0 \le h \le 50$ and $0 \le \rho \le 50$ with uniform probability density; at each point the initial direction in the source rest-frame was generated with an isotropic probability distribution. The photon trajectories were computed using the code developed by NZ08. 
For the disc-to-observer transfer we use the same transfer functions as in \texttt{reflkerr} and \texttt{reflkerr\_lp}.

Our model relies on the numerical accuracy of the code of NZ08, therefore, we now briefly compare it with other similar computations.
Irradiation of an accretion disc by a single point-like source, located off the symmetry axis of the Kerr metric, has been considered in a number of works, many of which aimed at explaining some observed variability effects by the change of the location of the source \citep[e.g.][NZ08]{1999ApJ...514..164R,2000MNRAS.315....1R,2004MNRAS.349.1435M,2010A&A...509A..22N}. There where some differences in the interpretation of the reduced variations of the reflected component in the model with a corotating source, which effect was attributed  to light bending by \citet{2004MNRAS.349.1435M}, whereas NZ08 clarified that it is mostly due to the assumed dependence of the source rotation velocity on its height. 
However, the quantitative results of these studies, e.g.\ the dependence of the reflection strength (typically represented by the Fe line flux in these works) on the source location, were in fairly good agreement.

\citet{2006A&A...453..773S} and NZ08 show radial profiles for disc irradiation, which can be compared to those presented by WF12, and again there seems to be a good agreement between these three works. In Figure \ref{fig:wf10} we show the emissivity profiles, computed with the code of NZ08, in the form suitable for the comparison with Figure 10 in WF12. 
We see that the profiles computed with these two codes match very well. The same was found for the case of the extended corona (Figure \ref{radial}(a)) and we note again that the difference between the conclusions of our and Wilkins et al.\  works do not result from any differences between the computational models, but from the different methods of their application to data interpretation.

\end{appendix}

\end{document}